\documentclass[superscriptaddress,aps,preprintnumbers,amsmath,amssymb,prd,nofootinbib,preprint]{revtex4-1}
\pdfoutput=1
\usepackage{graphicx}
\usepackage{epstopdf}
\usepackage{dcolumn}% Align table columns on decimal point
\usepackage{bm}% bold math
\usepackage{bbm}
\usepackage{hyperref}
\usepackage[usenames, dvipsnames]{color}
\usepackage{amsmath,amssymb}
\usepackage[sort&compress]{natbib}
\usepackage{float}
\usepackage{multirow}
\usepackage{slashed}
\usepackage{cancel}
\usepackage[table]{xcolor}
\usepackage{multirow}
\usepackage{booktabs}
\usepackage{makecell}
\usepackage{mathtools}
\usepackage{pifont}
\usepackage{enumitem}

\newcommand{\mpl}{M_{\rm Pl}}

\newcommand{\keV}{\textrm{ keV}}
\newcommand{\MeV}{\textrm{ MeV}}
\newcommand{\GeV}{\textrm{ GeV}}
\newcommand{\TeV}{\textrm{ TeV}}
\newcommand{\vev}[1]{\langle #1 \rangle}

\numberwithin{equation}{section}

\makeatletter
\renewcommand{\p@subsection}{}
\renewcommand{\p@subsubsection}{}
\makeatother

\makeatletter
\def\l@subsubsection#1#2{}
\makeatother

\hypersetup{colorlinks,citecolor= blue,linkcolor= black}

\begin{document}

\title{
Axiogenesis with a Heavy QCD Axion
}
\preprint{UMN-TH-4124/22, FTPI-MINN-22/15, CERN-TH-2022-090}

\author{Raymond T. Co}
\affiliation{\small William I. Fine Theoretical Physics Institute, University of Minnesota, Minneapolis, MN 55455, USA}
\affiliation{\small School of Physics and Astronomy, University of Minnesota, Minneapolis, MN 55455, USA}

\author{Tony Gherghetta}
\affiliation{\small School of Physics and Astronomy, University of Minnesota, Minneapolis, MN 55455, USA}

\author{Keisuke Harigaya}
\affiliation{\small Theoretical Physics Department, CERN, Geneva, Switzerland \\\vspace*{0.5 in}}

\begin{abstract}
We demonstrate that the observed cosmological excess of matter over antimatter may originate from a heavy QCD axion that solves the strong CP problem but has a mass much larger than that given by the Standard Model QCD strong dynamics. 
We investigate a rotation of the heavy QCD axion in field space, which is transferred into a baryon asymmetry through weak and strong sphaleron processes. This provides a strong cosmological motivation for heavy QCD axions, which are of high experimental interest. The viable parameter space has an axion mass $m_a$ between 1~MeV and 10 GeV and a decay constant $f_a < 10^5$ GeV, 
which can be probed by accelerator-based direct axion searches and observations of the cosmic microwave background. 
\end{abstract}

\maketitle

\begingroup
\hypersetup{linkcolor=black}
\tableofcontents
\endgroup

\newpage

\section{Introduction}

Strong CP violation in the Standard Model (SM) is due to the CP-odd $\theta$ term that arises from the nontrivial QCD vacuum structure. For non-zero quark masses, the non-perturbative $\theta$ parameter cannot be removed by chiral rotations of the quarks and therefore an $\mathcal{O}(1)$ $\theta$ parameter induces an electric dipole moment for the neutron~\cite{Bell:1969ts,Adler:1969gk,tHooft:1976rip} that is $\mathcal{O}(10^{10})$ times larger than the experimental constraint~\cite{Crewther:1979pi,Baker:2006ts}. This discrepancy is called the strong CP problem. The strong CP problem may be solved by the Peccei-Quinn (PQ) mechanism~\cite{Peccei:1977hh,Peccei:1977ur}, where a spontaneously broken $U(1)$ PQ symmetry and strong QCD dynamics yield a pseudo Nambu-Goldstone boson known as the QCD axion~\cite{Weinberg:1977ma,Wilczek:1977pj} that dynamically cancels the $\theta$ term.

The QCD axion can also play important cosmological roles. Recently, it was shown that in the early universe the QCD axion may undergo rotations in field space that occur even through the electroweak phase transition~\cite{Co:2019wyp}. The rotation corresponds to a non-zero PQ charge, and the charge is transferred into a quark chiral asymmetry by strong sphaleron processes and then further reprocessed into a baryon asymmetry by electroweak sphaleron processes~\cite{Co:2019wyp}. This mechanism is known as axiogenesis.

However, the minimal axiogenesis scenario faces a difficulty. The rotational kinetic energy eventually becomes the axion dark matter density~\cite{Co:2019jts}, and after requiring axiogenesis to reproduce the observed baryon asymmetry, the axion abundance exceeds the observed dark matter density. To resolve this overproduction of dark matter, several extensions of the axiogenesis scenario that enhance the baryon asymmetry have been proposed and their associated predictions have been investigated~\cite{Co:2019wyp,Co:2020jtv,Harigaya:2021txz,Co:2021qgl,Chakraborty:2021fkp,Kawamura:2021xpu}.

In this paper, we consider the production of the baryon asymmetry via axiogenesis with a QCD axion that obtains a large mass from non-QCD dynamics in a way that still solves the strong CP problem~\cite{Dimopoulos:1979pp, Rubakov:1997vp,Berezhiani:2000gh,Hook:2014cda,Fukuda:2015ana,Gherghetta:2016fhp,Agrawal:2017ksf, Agrawal:2017evu, Csaki:2019vte, Hook:2019qoh, Gherghetta:2020keg, Gherghetta:2020ofz}. Namely, the axion potential minimum remains nearly the same despite additional contributions to the axion potential from the non-QCD dynamics. With an enhanced mass, the heavy QCD axion can now decay well before the epoch of matter-radiation equality, so that the overproduction of dark matter in axiogenesis is avoided. Even though the heavy QCD axion is no longer a dark matter candidate, it changes roles and instead becomes responsible for generating the baryon asymmetry of the universe as well as solving the strong CP problem.

Such heavy QCD axions are motivated because they can more naturally explain the stringent requirements for the PQ symmetry. To solve the strong CP problem, the PQ symmetry must be preserved to a high quality, which means that the PQ symmetry is essentially only explicitly broken by the QCD anomaly. To illustrate the severity of this requirement, suppose there exists an explicit PQ symmetry breaking term $P^n/\mpl^{n-4}$, where $\mpl$ is the reduced Planck scale, and $P$ is the PQ symmetry breaking complex scalar field with a potential minimum at the scale $f_a$ known as the decay constant. In order for the axion potential minimum to solve the strong CP problem and not be displaced by more than $10^{-10}$, $n>8\mathchar`-36$ is required for $f_a=10^{8\mathchar`-16}$ GeV.  This clearly requires the global symmetry to be preserved to very high order beyond that expected from effective field theory.
These two aspects of the PQ symmetry, explicit breaking by the QCD anomaly and the extremely good symmetry of the perturbative potential, are best understood if the PQ symmetry accidentally arises from other exact symmetries~\cite{Georgi:1981pu}, in much the same way as how the baryon number symmetry of the Standard Model arises from gauge symmetry.

In particular, it is expected that quantum gravity explicitly breaks global symmetries~\cite{Perry:1978fd,Hawking:1979hw,Giddings:1988cx,Gilbert:1989nq,Kallosh:1995hi,Harlow:2018jwu,Harlow:2018tng} and therefore possibly generates higher-dimensional explicit PQ breaking terms. The exact symmetries, if gauged, can protect the PQ symmetry from quantum gravity effects~\cite{Holman:1992us,Barr:1992qq,Kamionkowski:1992mf,Dine:1992vx} and can be realized for the QCD axion but tend to require complicated symmetries (see~\cite{Lazarides:1985bj,Randall:1992ut,Dias:2002hz,Dias:2002gg,Babu:2002ic,Choi:2009jt,Carpenter:2009zs,Harigaya:2013vja,Harigaya:2015soa,Fukuda:2017ylt,DiLuzio:2017tjx,Fukuda:2018oco,Ibe:2018hir,Lillard:2018fdt} for additional literature along this direction).
Instead, if the QCD axion mass is larger than that obtained from QCD dynamics, the potential minimum is more stable against explicit breaking by higher dimensional operators. Furthermore, the decay constant can be much smaller than that allowed for the usual, light QCD axion (since the heavier axion mass evades supernova-cooling limits and beam-dump experiments), and thus the effect of higher dimensional operators is much more suppressed. 
For these two reasons, a simple exact symmetry can ensure a PQ symmetry of sufficiently high quality to solve the strong CP problem~\cite{Berezhiani:2000gh,Fukuda:2015ana,Fukuda:2017ywn}.

While the overproduction problem of dark matter is avoided, the paradigm of axiogenesis with a heavy QCD axion is still constrained by the following two requirements. Axion rotations must survive until after the electroweak phase transition, and the electroweak symmetry cannot be restored subsequently; otherwise the baryon asymmetry previously produced from axion rotations will be washed out by the sphaleron processes.
In the first requirement, the axion velocity must be larger than the mass to maintain its rotation. Moreover, when the axion field moves in an anharmonic potential, the axion rotation can fragment into axion fluctuations~\cite{Jaeckel:2016qjp, Berges:2019dgr, Fonseca:2019ypl,Morgante:2021bks} by parametric resonance~\cite{Dolgov:1989us, Traschen:1990sw, Kofman:1994rk, Shtanov:1994ce, Kofman:1997yn}. This axion fragmentation must therefore not occur before electroweak symmetry breaking. In the second requirement, the created axion fluctuations can thermalize via scattering with the thermal bath, which could reheat the universe and possibly restore electroweak symmetry.
These and other constraints are derived in detail for the following distinct cosmological histories of the heavy axion: when the universe undergoes the electroweak phase transition, 1) the axion rotates at the minimum of the PQ field potential, 2) the axion rotates on the body of the potential, or 3) parametric resonance occurs during this time. We find viable parameter space in the mass range 1 MeV - 10 GeV with a decay constant $f_a < 10^5 \GeV$, and interestingly much of this region can be probed by the collider and beam dump experiments as well as by observations of the cosmic microwave background (CMB).

The organization of the paper is as follows. In Sec.~\ref{sec:rotations}, we review the minimal axiogenesis scenario, from the initiation of the rotation to the production of the baryon asymmetry and dark matter. In Sec.~\ref{sec:models}, we review various models that generate a large axion mass consistent with the solution to the strong CP problem. In Sec.~\ref{sec:baryogenesis}, constraints of the mechanism are investigated for different cosmological scenarios and the viable parameter space is shown. Lastly, we conclude and summarize our findings in Sec.~\ref{sec:summary}, while Appendices~\ref{app:weak_KSVZ_model}, \ref{app:PR_numerics}, and \ref{app:wo} are dedicated to a KSVZ-type model construction, the numerical study of parametric resonance, and the analytical estimation of the washout rate of the rotation via the axion mass and the depletion of the radial mode, respectively.

%%%%%%%%%%%%%%%%%%%%%%%%%%%%
\section{Axion rotations and baryon asymmetry}
\label{sec:rotations}
%%%%%%%%%%%%%%%%%%%%%%%%%%%%
In this section, we review how axion rotations can be initiated from the dynamics of the PQ symmetry breaking field to produce the baryon asymmetry and the axion dark matter abundance in the context of the usual QCD axion.
We assume that the QCD axion arises from the angular component $\theta$ of a complex scalar field
\begin{align}
\label{eq:P}
    P = \frac{1}{\sqrt{2}}S e^{i \theta}~,
\end{align}
where $S$ is the radial component field whose vacuum expectation value is $f_a$, and we have assumed the domain wall number to be unity.

\subsection{Initiation and evolution of rotations}

The rotation of $P$ may be initiated in the same way as the Affleck-Dine  mechanism~\cite{Affleck:1984fy}.
The PQ symmetry is explicitly broken by a higher-dimensional potential term,
\begin{align}
    V = \frac{1}{M_{\rm UV}^{n-4}}P^n + {\rm h.c.},
    \label{eq:highdimpot}
\end{align}
where $M_{\rm UV}$ is a UV mass scale. Although such a term should be negligible near the vacuum value $S = f_a$, a non-negligible gradient to the angular component can initiate the rotation of $P$ if $S \gg f_a$ in the early universe. Such a large $S$ can be achieved if $S$ has a flat potential, which is natural in supersymmetric theories.
After the rotation begins, $S$ decreases by the cosmic expansion and the higher-dimensional term \eqref{eq:highdimpot} becomes negligible. The field $P$ continues to rotate conserving angular momentum, which is nothing but the PQ charge associated with the PQ symmetry,
\begin{align}
    n_{\rm PQ} = \dot{\theta} S^2.
\end{align}
It is also possible to initiate the axion rotation by transferring the charge of another rotating scalar field into the axion~\cite{Domcke:2022wpb}, with the rotating scalar field initiated by the Affleck-Dine mechanism. In this case, the potential of $S$ does not have to be flat.

Initially the rotation is not circular and has non-zero ellipticity; it includes both angular and radial motion. The field $P$ can be thermalized via its interaction with the thermal bath. The radial motion dissipates, while the angular motion remains because of PQ charge conservation. The PQ charge can be converted into a quark chiral asymmetry in the thermal bath via the strong sphaleron process, but it is free-energetically favored to keep almost all charges in the form of rotation, with a small fraction $\sim T^2/S^2$ converted to the chiral asymmetry, as long as the radius of the rotation is larger than the temperature~\cite{Laine:1998rg,Co:2019wyp}. The resultant motion after thermalization is circular without any ellipticity.

The explicit breaking of quark chiral asymmetries leads to a slow washout of the PQ charge. If one quark chiral asymmetry is unbroken, a linear combination of the PQ symmetry and the quark chiral asymmetry remains exact and the washout does not occur. Therefore, the washout rate is suppressed by the smallest Yukawa coupling, namely the up quark Yukawa coupling, $y_u$. Taking into account the small fraction of the quark chiral asymmetry $\sim T^2/S^2$ in comparison with the PQ charge in the rotation, the washout rate is
\begin{align}
    \Gamma_{{\rm wo}, u} \simeq \alpha_s y_u^2 \frac{T^3}{S^2},
\end{align}
where $\alpha_s=g_s^2/4\pi$ is the QCD coupling.
For the usual QCD axion with $f_a > 10^8$ GeV, the washout rate does not exceed the Hubble expansion rate and the axion rotation is not washed out.
However, for the heavy QCD axion with low $f_a$, avoiding the washout gives constraints on the parameter space.

\subsection{Baryon asymmetry}

The PQ charge is transferred into a baryon asymmetry in the following way, known as axiogenesis~\cite{Co:2019wyp}.
The PQ symmetry and quark chiral symmetries have a QCD anomaly, so the strong sphaleron process transfers the PQ charge into quark chiral charges. The quark chiral symmetry and the $B+L$ symmetry have an electroweak anomaly, so the weak sphaleron process transfers the quark chiral charges into a $B+L$ asymmetry.
These processes are in thermal equilibrium and the baryon asymmetry at the temperature $T$ is given by
\begin{align}
    n_B = c_B\, \dot{\theta}(T) T^2,
\end{align}
where $c_B\sim 0.1$ is a model-dependent coefficient that parameterizes the efficiency of the charge transfer~\cite{Co:2019wyp}.

The weak sphaleron process freezes out near the electroweak phase transition when the sphaleron rate falls below the Hubble rate.
The baryon asymmetry is also frozen at this temperature, and the baryon asymmetry normalized by the entropy density $s$ is given by
\begin{align}
\label{eq:YB}
    \frac{n_B}{s} = \frac{c_B \dot{\theta} T^2}{\frac{2\pi^2}{45} g_* T^3}\Bigg|_{T = T_{\rm ws}} =~& 8.7\times10^{-11}\left(\frac{\dot{\theta}(T_{\rm ws})}{5.3\keV}\right) \left(\frac{130\GeV}{T_{\rm ws}}\right) \left(\frac{c_B}{0.1}\right), \nonumber \\
    =~& 8.7\times10^{-11} \left(\frac{Y_{\rm PQ}}{510}\right) \left(\frac{c_B}{0.1}\right) \left(\frac{10^8\GeV}{f_a}\right)^2 \left(\frac{T_{\rm ws}}{130\GeV}\right)^2,
\end{align}
where $g_*$ is the number of Standard Model degrees of freedom in the thermal bath, $T_{\rm ws}$ is the temperature below which the weak sphaleron process becomes ineffective, and $Y_{\rm PQ}=n_{\rm PQ}/s$. The observed baryon asymmetry yield is $Y_B^{\rm obs} \equiv n_B^{\rm obs}/s = 8.7 \times 10^{-11}$~\cite{Aghanim:2018eyx}.

\subsection{Overproduction of axion dark matter}
\label{sec:DM}

For the conventional, light QCD axion, the rotation can be rapid enough and continue to a low temperature where the axion oscillation would occur~\cite{Co:2019jts}. In this case, the kinetic energy of the axion rotation is converted to an axion dark matter density in a process called the kinetic misalignment mechanism. In Ref.~\cite{Co:2019jts}, the effect of the delay of the axion field zero-mode oscillation around the axion potential minimum was considered, where it was found that the axion number density is $n_a \simeq n_{\rm PQ}$. 

However, as discussed in Ref.~\cite{Fonseca:2019ypl} in the context of the relaxion (see also Refs.~\cite{Jaeckel:2016qjp, Berges:2019dgr}), non-zero momentum modes can be produced via parametric resonance~\cite{Dolgov:1989us,Traschen:1990sw,Kofman:1994rk,Shtanov:1994ce,Kofman:1997yn}. Neglecting the Hubble expansion, the equation of motion of a non-zero mode $a_k$ is given by
\begin{align}
  \ddot{a}_k + \left[k^2 + m_a^2 {\rm cos}^2(\dot{\theta}t)\right]a_k =0.  
\end{align}
This differential equation can be recast into the Mathieu equation. For $\dot{\theta} \gg m_a$, the resonance is narrow and the first resonance band occurs at
\begin{align}
    k_{\rm peak} = \frac{\dot{\theta}}{2},\qquad\frac{\Delta k}{k} \simeq \frac{m_a^2}{\dot{\theta}^2},
\end{align}
with the growth rate at the peak given by
\begin{align}
\label{eq:PR}
    \Gamma_{\rm PR} \simeq \frac{m_a^2}{\dot{\theta}}~.
\end{align}

Once this parametric resonance becomes effective, the zero-mode rotation fragments into axion fluctuations.
The narrow resonance causes the following two effects to suppress the fragmentation: 1) production of axion fluctuations that decreases $\dot{\theta}$ and shifts the resonance band~\cite{Fonseca:2019ypl} and 2) cosmic expansion that decreases $\dot{\theta}$ and $k$~\cite{Co:2021lkc}. As a result, the effective parametric resonance rate is smaller than that in Eq.~(\ref{eq:PR})~\cite{Fonseca:2019ypl},
\begin{align}
\label{eq:PReff}
    \Gamma_{\rm PR,eff} \simeq \frac{m_a^4}{\dot{\theta}^3}.
\end{align}

The number density of axions produced from the parametric resonance is given by~\cite{Co:2021rhi}
\begin{align}
    n_a \simeq \frac{\rho_{\rm rot}}{k_{\rm peak}} =\frac{\dot{\theta}^2 f_a^2 /2}{\dot{\theta}/2} = \dot{\theta} f_a^2 = n_{\rm PQ},
\end{align}
so the axion number density is as large the PQ charge density and gives the same result as the zero-mode analysis in Ref.~\cite{Co:2019jts}.

For the QCD axion with a PQ charge yield that explains the baryon asymmetry of the universe, the angular velocity around the QCD phase transition is indeed sufficiently large that kinetic misalignment is at work for the light QCD axion. The resultant axion energy density normalized by the observed dark matter density is
\begin{align}
    \frac{\rho_a}{\rho_{\rm DM}} \simeq 70 \left( \frac{f_a}{10^8~{\rm GeV}} \right) \left( \frac{130~{\rm GeV}}{T_{\rm ws}} \right)^2 \left( \frac{0.1}{c_B} \right).
\end{align}
One can see that for $f_a$ satisfying the astrophysical lower bound $f_a >10^8$ GeV~\cite{Chang:2018rso}, axion dark matter is overproduced by the kinetic misalignment mechanism. This problem cannot be avoided by entropy production, since the baryon asymmetry is also diluted.

Instead, we will consider a heavy QCD axion so that the axions produced by the kinetic misalignment mechanism can decay, avoiding the overproduction problem. Thus, in our scenario the axion will become responsible for generating the baryon asymmetry while simultaneously solving the strong CP problem (with a high quality).

%%%%%%%%%%%%%%%%%%%%%%%%%%%%
\section{Heavy QCD axion models}
\label{sec:models}
%%%%%%%%%%%%%%%%%%%%%%%%%%%%
In this section, we review models of generating a heavy QCD axion.
As discussed in the introduction, such a scenario can more easily explain the origin of the PQ symmetry.

%%%%%%%%%%%%%%%%%%%%%%%%%%%%
\subsection{UV 4D instanton}
\label{subsec:UV_4D}
%%%%%%%%%%%%%%%%%%%%%%%%%%%%

If $SU(3)_c$ is extended to be the diagonal subgroup of a parent product group $SU(3)^k=SU(3)_1\times SU(3)_2\times \dots \times SU(3)_k$ that is spontaneously broken to $SU(3)_c$ at some UV scale $M$, then UV instantons can generate a QCD axion mass that is larger than the usual IR contribution at the QCD scale~\cite{Agrawal:2017ksf, Csaki:2019vte}. In this setup, $k$ axions $a_i$ $(i=1\dots k)$ are introduced and couple to the $k$ $SU(3)$ $G{\widetilde G}$ terms, thereby eliminating the $k$ theta terms. Furthermore, the Standard Model quarks are assumed to be charged under only one of the parent $SU(3)$ groups, $SU(3)_1$, with no additional colored fermions charged under the other $SU(3)$ groups. 

First, suppose that $f_a$ is larger than the inverse size of the UV instantons. Then we may integrate out the KSVZ quarks and use the dimension-five coupling between the axion and the gluon.
The axion mass generated from the instantons at the energy scale $\mu$ is
\begin{align}
\label{eq:mass_4D}
    m_{a}^2 \sim \kappa_q\, \frac{\mu^4}{f_a^2}\,e^{- \frac{2\pi}{\alpha_s (\mu)}},
\end{align}
where $\alpha_s (\mu)$ is the QCD coupling and $\kappa_q\sim {\cal O}(10^{-23})$ is the suppression from the SM quark chiral symmetries.
In the SM QCD, the factor $\mu^4 e^{- \frac{2\pi}{\alpha_s (\mu)}}$ is a decreasing function of $\mu$ due to asymptotic freedom, so that UV instantons are never important. However, when $SU(3)^k$ is broken to $SU(3)_c$ at the symmetry breaking scale $M$, the QCD gauge coupling is obtained from the matching condition 
\begin{align}
    \frac{1}{\alpha_s(M)} = \sum_{i=1}^k \frac{1}{\alpha_i(M)}.
\end{align}
This implies that each individual coupling $\alpha_i(\mu)$ must be larger than the SM QCD value $\alpha_s(\mu)$ and therefore the enhanced couplings of the individual $SU(3)$ groups can make the UV instanton effects dominate over the IR instanton effect.

The small instanton contributions can enhance the axion mass over the QCD contributions for very high breaking scales $M$ and $k\geq 3$~\cite{Agrawal:2017ksf}. A full calculation~\cite{Csaki:2019vte} that directly computes the Higgs loop effects obtains for $k=3$ and $M=10^{14}$ GeV a maximum enhancement of $4\times 10^{10}$ for the (lightest) axion mass compared to the QCD contribution. In the limit $k\gg1$, the axion masses scale as $m_{a_1}\sim \sqrt{\kappa_q} M^2/f_{a_1}$ and $m_{a_i}\sim M^2/f_{a_i}$ for $i=2\dots k$, showing that the lightest axion mass is much heavier than the QCD axion mass contribution for $M\gg \Lambda_{\rm QCD}$, where $\Lambda_{\rm QCD}$ is the QCD strong coupling scale.

Instead, if $f_a < M$, one must UV complete the dimension-five axion-gluon coupling and, for example, include the contribution of KSVZ quarks to compute the axion mass enhancement. For simplicity, we will consider one pair of KSVZ quarks ($Q,{\bar Q}$) in the fundamental representation and include the KSVZ quarks in the t' Hooft vertex at the scale $M$. Assuming ${\cal L}_Q=y_Q P {\bar Q}Q +{\rm h.c.}$, the KSVZ quark legs are closed using this Yukawa interaction which generates a tadpole term for $P$, 
\begin{align}
    \kappa_q y_Q P M^3 e^{- \frac{2\pi}{\alpha_1(M)}} + {\rm h.c.} .
\end{align}
This tadpole term gives rise to an axion mass which is suppressed by  
$m_Q \sim y_Q S < M$ compared to the case with $m_Q > M$. On the other hand, the 
QCD coupling at the scale $M$ is larger than the case with $m_Q > M$ because of the KSVZ quark contribution to the running QCD coupling, giving a smaller action which enhances 
the axion mass. The net result of these two effects causes an extra axion mass suppression given by
\begin{align}
\label{eq:axionsup}
m_{a}\Big|_{m_Q < M} \simeq  \left(\frac{m_Q}{M}\right)^{1/6} \times m_a\Big|_{m_Q> M}.
\end{align}
For example, assuming $m_Q = 10^4$ GeV and $M = 10^{14}$ GeV, the extra suppression factor is ${\cal O}(10^{-2})$. 

Using these results, we can determine the 4D instanton parameter values that give axion masses in the range $1\, {\rm MeV}\lesssim m_{a_1} \lesssim 10\, {\rm GeV}$, which will be relevant for the axiogenesis mechanism discussed in Sec.~\ref{sec:baryogenesis}. For example, assuming $f_a\sim 10^4$ GeV, the usual QCD contribution is $m_a^{\rm QCD}= 570$~eV~\cite{GrillidiCortona:2015jxo} and requires an axion mass enhancement of $\sim 10^4$ to obtain $m_{a_1}\sim 10$ MeV. Given that $f_a <M$, and therefore including the extra suppression \eqref{eq:axionsup}, the required axion masses can be obtained for $k=3$ and $M \sim 10^{8}$ GeV~\cite{Csaki:2019vte}. Similarly, axion masses $\gtrsim 1$ GeV with 10 GeV $\lesssim f_a\lesssim 10^3$ GeV are obtained for $M\sim 10^9$ GeV. However, to avoid light KSVZ quarks requires going beyond the minimal KSVZ model and a particular model is given in Appendix~\ref{app:weak_KSVZ_model}.

Furthermore, in the axiogenesis scenario, the field value of $P$ in the early universe is different from the vacuum value, and thus one needs to know the axion mass at high temperatures or equivalently its dependence on $S$, as discussed in Sec.~\ref{sec:DM} in the context of parametric resonance. For example, when one pair of KSVZ quarks is present, the axion mass squared, $m_a^2 \propto S^{-1}$. In general, for $n_Q$ pairs of KSVZ quarks, a potential term $P^{n_Q}$ is induced by the UV 4D instantons, giving $m_a^2 \propto S^{n_Q-2}$, and therefore unless $n_Q=2$, the field-dependence of the axion mass must be taken into account to study the dynamics of $P$.

%%%%%%%%%%%%%%%%%%%%%%%%%%%%
\subsection{UV 5D instanton}
\label{subsec:5D}
%%%%%%%%%%%%%%%%%%%%%%%%%%%%

An alternative possibility to enhance the axion mass from UV small instantons is to assume that QCD gluons propagate in a 5th dimension at high energies~\cite{Gherghetta:2020keg}. It is well known that above the compactification scale $1/R$ the QCD coupling $\alpha_s$ becomes large again, therefore giving rise to a new UV instanton contribution to the axion mass.

A minimal setup is to assume a flat extra dimension compactified on a $Z_2$ orbifold with the SM quarks confined to one of the boundaries. The enhancement of the axion mass follows from the fact that there is now a power-law term $R/\rho$, where $\rho$ is the instanton size, in the effective action in the instanton density that arises from the positive frequency modes of the Kaluza-Klein gluon states in the instanton background. This leads to the axion mass squared~\cite{Gherghetta:2020keg}
\begin{equation}
\label{eq:mass_5D}
    m_{a}^2 \sim \kappa_q\, \frac{1}{f_a^2 R^4}\,e^{- \frac{2\pi}{\alpha_s (1/R)}+\Lambda_5 R},
\end{equation}
where $\Lambda_5$ is the 5D strong coupling scale. 
Note that the maximum possible enhancement for the axion mass occurs when the 5D theory is strongly coupled at $\Lambda_5$ leading to the naive dimensional analysis estimate $m_a^2\sim \kappa_q \Lambda_5^4/f_a^2$. 

To implement the axiogenesis mechanism with a large angular velocity, it is crucial that the initial field value of the PQ symmetry breaking field $ S\gg f_a$. In the minimal setup the axion is identified as the 5th component of a bulk $U(1)$ gauge field with 
$f_a\sim 1/R$ that would lead to field values $ S \gg 1/R$ and require a 5D description. To maintain an effective 4D description of the axiogenesis mechanism, the minimal setup can be modified to assume that the axion is a localized boundary field which couples to $G{\widetilde G}$. 
The decay constant $f_a$ can now be generated by adding a boundary potential and KSVZ quarks that induce a linear $P$ term via the 5D instanton (similar to the 4D instanton case). The decay constant is then a free parameter and can be chosen to satisfy $f_a \ll  S  < 1/R$. Constraints from possible higher-dimensional CP breaking terms on the boundaries are given in~\cite{Bedi:2022qrd}.

For example, assuming that only one pair of KSVZ quarks is localized on the UV boundary, a tadpole term for $P$ can be generated analogous to the 4D instanton case. Since $f_a< 1/R$, there is an extra suppression in the axion mass enhancement from the KSVZ quarks given by
\begin{align}
\label{eq:axionsup5D}
m_{a}\Big|_{m_Q < 1/R} \simeq  \left(m_Q R\right)^{1/6} \times m_a\Big|_{m_Q>1/R} \, ,
\end{align}
where $m_a\Big|_{m_Q>1/R}$ is given by Eq.~\eqref{eq:mass_5D} with $\alpha_s(1/R)$ determined by the running without the KSVZ quarks.
To obtain ${\cal O}({\rm 10\, MeV})$ axion masses with $f_a\sim 10^4$ GeV corresponding to an enhancement of order $10^{4}$ over the usual QCD axion mass requires $1/R \sim 10^8$~GeV for perturbativity parameter values $\epsilon \sim 0.3$ where the perturbativity parameter $\epsilon$ is defined via the relation, $\Lambda_5 R= 6\pi \epsilon/\alpha_s(1/R)$~\cite{Gherghetta:2020keg}. Similarly, allowed axion masses with 10 GeV $\lesssim f_a \lesssim 10^3$ GeV require $1/R\sim 10^9$ GeV assuming $\epsilon \sim 0.3$.

Finally, as in the case with UV 4D instantons, the field-dependent axion mass must be taken into account unless $n_Q=2$ in order to study the dynamics of $P$.

%%%%%%%%%%%%%%%%%%%%%%%%%%%%
\subsection{Mirror QCD}
\label{subsec:mirror}
%%%%%%%%%%%%%%%%%%%%%%%%%%%%
Next we consider a mirror copy of the SM and a $Z_2$ symmetry that exchanges the SM with the mirror SM. The QCD axion is assumed to be $Z_2$ neutral and couples to both QCD and mirror QCD,
\begin{align}
    \frac{1}{64\pi^2}\frac{a}{f_a}\varepsilon^{\mu\nu \rho \sigma}\left(G_{\mu\nu} G_{\rho \sigma } + G'_{\mu\nu} G'_{\rho \sigma }\right),
\end{align}
where $G_{\mu\nu} (G'_{\mu\nu})$ is the QCD (mirror QCD) field strength.
Because of the $Z_2$ symmetry, the theta terms as well as the couplings of QCD and mirror QCD to the axion are the same. Even though the axion obtains an extra mass from the mirror QCD dynamics, the strong CP problem is still solved~\cite{Rubakov:1997vp,Berezhiani:2000gh,Hook:2014cda,Fukuda:2015ana, Hook:2019qoh}.

The axion mass contribution from mirror QCD may be much larger than the usual QCD contribution. This is achieved by requiring a larger mirror electroweak scale VEV $v'$ than the electroweak scale VEV $v$ that makes the mirror quarks heavier than the SM quarks, thereby accelerating the running of the gauge coupling toward the IR. Assuming $v'\gg v$, this $Z_2$ breaking can arise spontaneously from radiative corrections to the Higgs and mirror Higgs potential~\cite{Hall:2018let,Dunsky:2019upk} or from a $Z_2$ odd order parameter that couples to the Higgs and mirror Higgs field.
For KSVZ-type models~\cite{Fukuda:2015ana}, the axion mass is then given by
\begin{align}
    m_a \simeq 40 \MeV \left(\frac{v'}{10^9\GeV}\right)^{8/11} \left(\frac{10^5 \GeV}{f_a}\right),
\end{align}
where the mirror quark masses are assumed to be above the mirror QCD scale.
For DFSZ-type models~\cite{Berezhiani:2000gh}, $f_a\sim v'$ and the axion mass is $\lesssim$ MeV. Also, constraints on the decay constant from beam-dump experiments are stronger.

%%%%%%%%%%%%%%%%%%%%%%%%%%%%
\section{Baryon asymmetry from a heavy QCD axion}
\label{sec:baryogenesis}
%%%%%%%%%%%%%%%%%%%%%%%%%%%%

In this section, we derive the conditions for the baryon asymmetry to be explained by the axion rotations via axiogenesis elaborated in Sec.~\ref{sec:rotations}. We only assume that the axion has interactions with the gluons and a mass larger than that provided by pure QCD dynamics, which can be achieved by the models in Sec.~\ref{sec:models}. There remain distinct possibilities of cosmological evolution because, during the electroweak phase transition, the axion rotation may already
be at the minimum of the PQ field potential as we will study in Sec.~\ref{sec:const_mass} or else the axion rotation will be on the body of potential as will be studied in Sec.~\ref{sec:varying_mass}.
The implications for a mirror QCD model will be discussed in Sec.~\ref{sec:mirror_QCD}. Lastly, Sec.~\ref{sec:PR_EW} analyzes the special case where parametric resonance and the electroweak phase transition occur simultaneously.

\subsection{Constant axion mass}
\label{sec:const_mass}
We first consider the case where the axion mass is constant in temperature and the radial component is already at the PQ potential minimum at the electroweak phase transition. If there is no entropy production after the electroweak phase transition, the angular velocity of the axion when the electroweak sphaleron processes fall out of equilibrium must be
\begin{align}
\label{eq:dtheta_SM}
    \dot\theta(T_{\rm ws})  = \frac{2 \pi^2}{45} g_*(T_{\rm ws}) \frac{ T_{\rm ws} Y_B^{\rm obs}}{c_B} 
     \simeq 5 \keV \left( \frac{T_{\rm ws}}{130 \GeV} \right) \left( \frac{0.1}{c_B} \right) \left( \frac{g_*(T_{\rm ws})}{106.75} \right) ,
\end{align}
based on Eq.~(\ref{eq:YB}). Here $T_{\rm ws} = 130 \GeV$ is the prediction of the Standard Model~\cite{DOnofrio:2014rug}. For consistency with the assumption that the axion is rotating at $T_{\rm ws}$, the kinetic energy must be larger than the potential barriers, i.e., $\dot\theta > 2 m_a$. However, the BBN constraint $m_a \gtrsim {\rm MeV}$ implies an overproduction of the baryon asymmetry according to Eq.~(\ref{eq:dtheta_SM}). Therefore, in what follows, we consider the scenario where the universe undergoes a period of reheating during the electroweak phase transition by either the inflaton or generic moduli so that the baryon asymmetry is diluted. 

\begin{figure}[!t]
    \centering
    \includegraphics[width=0.495\columnwidth]{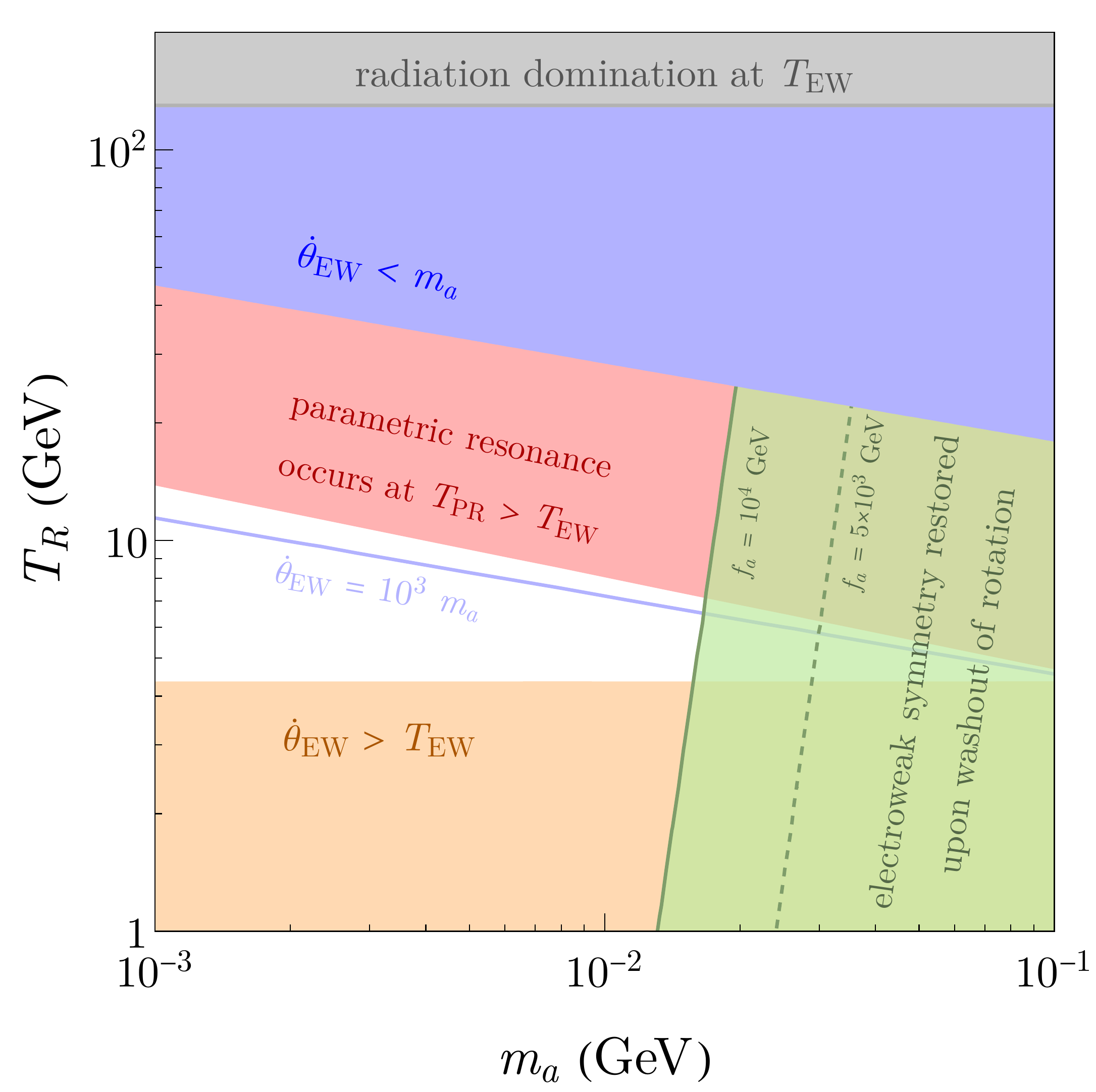}
    \includegraphics[width=0.495\columnwidth]{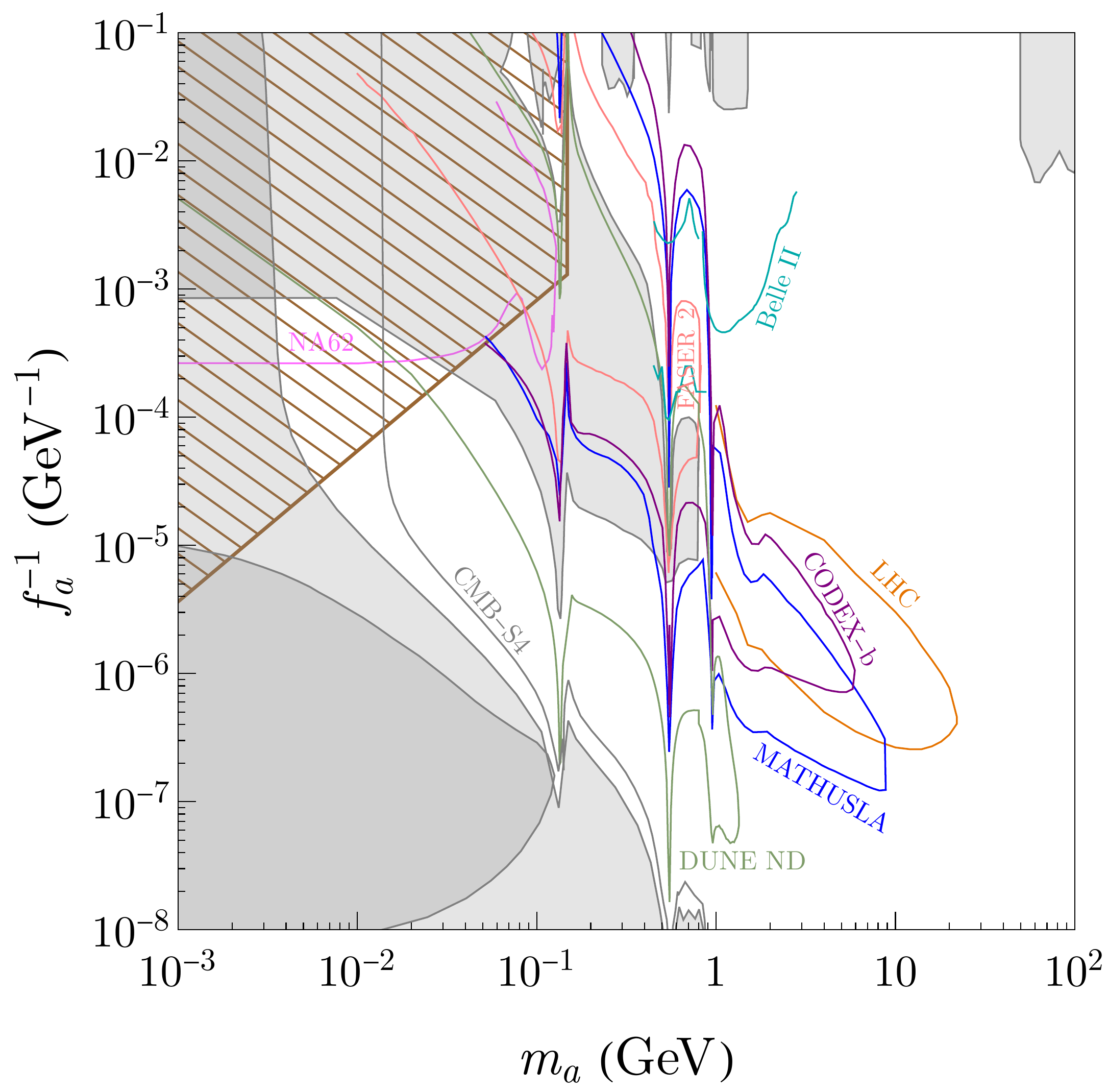}
    \caption{{\bf Left:} Individual constraints that determine the viable parameter space in the right panel are explicitly shown with the reheat temperature $T_R$ as a free parameter. {\bf Right:} The brown hatched region shows the viable parameter space where the baryon asymmetry can be explained by axiogenesis with a heavy QCD axion, specifically for the case where the axion mass is a constant after the electroweak phase transition.}
    \label{fig:ConstMass}
\end{figure}

For reheating with a constant rate, e.g., via perturbative decays, the temperature scales as $T \propto a^{-3/8}$ when the entropy is efficiently produced from reheating. For maximal dilution, we focus on the case where entropy is already efficiently produced at $T = T_{\rm ws}$. Accordingly, the baryon asymmetry yield $Y_B = n_B / s$ scales as $a^{-3}/T^3 \propto T^5$ until the end of reheating at $T=T_R$. Equivalently, this means that the baryon asymmetry is diluted by a factor of $(T_{\rm ws}/ T_R)^5$. The required angular velocity is then
\begin{align}
\label{eq:thetadot_TR}
    \dot\theta(T_{\rm ws}) & \simeq 2 \GeV 
    \left( \frac{T_{\rm ws}}{130 \GeV} \right)^6
    \left( \frac{0.1}{c_B} \right) 
    \left( \frac{g_*(T_{\rm ws})}{106.75} \right) 
    \left( \frac{10 \GeV}{T_R} \right)^5 .
\end{align}

We now discuss several constraints on the reheat temperature $T_R$. The aforementioned constraint $\dot\theta(T_{\rm ws}) > m_a$ is satisfied for
\begin{align}
\label{eq:TRmax_thetadot_ma}
    T_R \lesssim 50 \GeV 
    \left(\frac{T_{\rm ws}}{130 \GeV} \right)^{ \scalebox{1.01}{$\frac{6}{5}$} } 
    \left( \frac{0.1}{c_B} \right)^{ \scalebox{1.01}{$\frac{1}{5}$} }
    \left( \frac{\rm MeV}{m_a} \right)^{ \scalebox{1.01}{$\frac{1}{5}$} }.
\end{align}
The blue region in the left panel of Fig.~\ref{fig:ConstMass} is therefore excluded. For reference, the blue line shows the contour when $\dot\theta(T_{\rm ws}) = 10^3 m_a$.
On the other hand, in the derivation of the baryon asymmetry, it is assumed that $\dot\theta(T_{\rm ws}) < T_{\rm ws}$, which is satisfied when%
\footnote{
The following happens when this condition is violated. Given  that $\dot{\theta} >T$ and $S>T$ imply the rotation dominates the energy density of the universe, the potential of $S$ must then be quadratic and thus $\dot{\theta}$ remains constant when $S > f_a$~\cite{Co:2019wyp}.
Although $\dot{\theta} < T$ as the rotation is initiated, $T$ eventually falls below $\dot\theta$.
The Higgs field obtains a large chemical potential $\sim \dot{\theta}$ and is destabilized to obtain a field value $\sim \dot{\theta} > T$, so the sphaleron process becomes ineffective. Once $S$ reaches $f_a$, $\dot{\theta}$ decreases rapidly and becomes smaller than $T$ after $T=T_{\rm ws}$. The baryon asymmetry is fixed when $\dot{\theta}$ first becomes larger than $T$. One can show that the resultant baryon asymmetry is smaller than that for $\dot{\theta}(T_{\rm ws})\sim T_{\rm ws}$. Hence, the baryon asymmetry is underproduced in the orange shaded region.
}
\begin{align}
\label{eq:TRmin_thetadot_TEW}
    T_R \gtrsim 4 \GeV \left(\frac{T_{\rm ws}}{130 \GeV} \right) \left( \frac{0.1}{c_B} \right)^{ \scalebox{1.01}{$\frac{1}{5}$} }.
\end{align}
This excludes the orange region in the left panel of Fig.~\ref{fig:ConstMass}.
Since $\dot\theta$ originates from the mass of the radial component, which in turn must be smaller than $f_a$, we must impose $\dot\theta(T_{\rm ws}) < f_a$, leading to a lower bound on the reheat temperature
\begin{align}
\label{eq:TRmin_thetadot_fa}
    T_R \gtrsim 2 \GeV 
    \left(\frac{T_{\rm ws}}{130 \GeV} \right)^{ \scalebox{1.01}{$\frac{6}{5}$} } 
    \left( \frac{0.1}{c_B} \right)^{ \scalebox{1.01}{$\frac{1}{5}$} }
    \left( \frac{10^4 \GeV}{f_a} \right)^{ \scalebox{1.01}{$\frac{1}{5}$} }.
\end{align}

While the axion is moving over the potential barriers, the axion self-interactions lead to parametric resonance~\cite{Dolgov:1989us, Traschen:1990sw, Kofman:1994rk, Shtanov:1994ce, Kofman:1997yn}, and the coherent rotation fragments into axion fluctuations~\cite{Jaeckel:2016qjp, Berges:2019dgr, Fonseca:2019ypl,Morgante:2021bks}. The growth rate of the axion fluctuations is given by Eq.~(\ref{eq:PReff}).
Using the required $\dot\theta(T_{\rm ws})$ from Eq.~(\ref{eq:thetadot_TR}) and the scaling $\dot\theta \propto a^{-3} \propto T^8$, one finds that parametric resonance occurs when $\Gamma_{\rm PR} \simeq 10 H$ at a temperature
\begin{align}
\label{eq:TPR}
    T_{\rm PR} \simeq 100 \GeV 
    \left( \frac{T_{\rm ws}}{130 \GeV} \right)^{ \scalebox{1.01}{$\frac{3}{14}$} }
    \left( \frac{c_B}{0.1} \right)^{ \scalebox{1.01}{$\frac{3}{28}$} }
    \left( \frac{T_R}{10 \GeV} \right)^{ \scalebox{1.01}{$\frac{17}{28}$} }
    \left( \frac{m_a}{\rm MeV} \right)^{ \scalebox{1.01}{$\frac{1}{7}$} } 
    \left( \frac{106.75}{g_*(T_{\rm ws})} \right)^{ \scalebox{1.01}{$\frac{1}{8}$} } .
\end{align}
For successful baryogenesis to occur by the coherent axion rotation, parametric resonance should not occur before $T = T_{\rm ws}$, giving an upper bound on the reheat temperature
\begin{align}
    T_R \lesssim 10 \GeV 
    \left(\frac{T_{\rm ws}}{130 \GeV} \right)^{ \scalebox{1.01}{$\frac{22}{17}$} } 
    \left( \frac{0.1}{c_B} \right)^{ \scalebox{1.01}{$\frac{3}{17}$} }
    \left( \frac{g_*(T_{\rm ws})}{106.75} \right)^{ \scalebox{1.01}{$\frac{7}{34}$} }
    \left( \frac{\rm MeV}{m_a} \right)^{ \scalebox{1.01}{$\frac{4}{17}$} }.
\end{align}
This constraint excludes the red region in the left panel of Fig.~\ref{fig:ConstMass}.

The axion fluctuations created from parametric resonance scatter with gluons to produce radiation that can reheat the universe. The rate of axion-gluon scattering is given by
\begin{align}
\label{eq:agg_rate}
    \Gamma_{agg} \simeq \frac{1}{4\pi} \left( \frac{\alpha_s}{4\pi} \right)^2 \frac{k_a^2 T}{f_a^2},
\end{align}
with $k_a$ the energy of each axion, which is of order $\dot\theta(T_{\rm PR})$.
The universe should not be reheated above the electroweak scale, which would restore the electroweak symmetry and the baryon asymmetry produced previously will be washed out by the sphaleron processes. This constraint is expressed as
\begin{align}
\label{eq:rho_rot_TEW}
    \rho_{\rm rot} \times \min\left(1, \frac{\Gamma_{agg}}{H}\right) < \frac{\pi^2}{30}  g_*(T_{\rm ws}) T_{\rm ws}^4 .
\end{align}
In the parameter range of interest, $\Gamma_{agg}>H$ is always true at $T = T_{\rm PR}$, in which case Eq.~(\ref{eq:rho_rot_TEW}) leads to the constraint 
\begin{align}
    T_R \gtrsim 10 \GeV 
    \left(\frac{130 \GeV}{T_{\rm ws}} \right)^{16} 
    \left( \frac{0.1}{c_B} \right)
    \left( \frac{m_a}{10 \MeV} \right)^8
    \left( \frac{f_a}{10^4 \GeV} \right)^7
    \left( \frac{106.75}{g_*(T_{\rm ws})} \right)^{ \scalebox{1.01}{$\frac{7}{2}$} }.
\end{align}
This constraint is shown in Fig.~\ref{fig:ConstMass} as the green region, where the solid (dashed) boundary assumes $f_a = 10^4 \GeV$ ($f_a = 5\times 10^3 \GeV$). Therefore, for a given $f_a$, there is an upper bound on $m_a$ due to the green and red regions. The viable parameter space in $m_a$ and $f_a$ is accordingly obtained and displayed as the brown hatched region in the right panel of Fig.~\ref{fig:ConstMass}. The brown lower boundary is understood from the inconsistency between the green and red regions in the left panel, while the brown right boundary is due to the incompatibility between the red and orange regions in the left panel.
The gray-shaded regions are excluded by the dark radiation constraints from the CMB~\cite{Dunsky:2022uoq}, the supernova-cooling bound~\cite{Chang:2018rso}, or accelerator searches~\cite{Dobrich:2015jyk,Dolan:2017osp,Aloni:2018vki,FASER:2018eoc,Ertas:2020xcc,NA64:2020qwq,Gori:2020xvq}. Much of the predicted parameter space can be probed by the kaon decay searches at NA62~\cite{Ertas:2020xcc} and future CMB observations~\cite{Dunsky:2022uoq}, whose sensitivities are shown by the magenta and gray lines, respectively. We include the sensitivities of accelerator searches at the HL-LHC~\cite{Hook:2019qoh} in orange, CODEX-b~\cite{Gligorov:2017nwh} in purple, MATHUSLA~\cite{Chou:2016lxi} in blue, FASER 2~\cite{Feng:2018pew,FASER:2018eoc} in pink, Belle II~\cite{Chakraborty:2021wda,Bertholet:2021hjl} in cyan, and DUNE near detector~\cite{Kelly:2020dda} in green.

\subsection{Axion mass dependence on radial field value}
\label{sec:varying_mass}

In this section, we consider the possibility that the axion mass at the electroweak scale is different from the vacuum value of the mass, which is in contrast to the constant mass scenario analyzed in Sec.~\ref{sec:const_mass}. A varying axion mass can be the case if the explicit PQ breaking term depends on the radial field value $S$ and the field has not been relaxed to the minimum $f_a$ by $T_{\rm ws}$, i.e., $S(T_{\rm ws}) > f_a$. The field-value dependence of the axion mass appears in the UV models presented in Sec.~\ref{sec:models}.

The required angular velocity at the electroweak phase transition $\dot\theta(T_{\rm ws})$ remains the same as Eq.~(\ref{eq:thetadot_TR}). Several bounds considered in Sec.~\ref{sec:const_mass} also identically apply in this scenario because the axion mass and/or the radial field value at $T_{\rm ws}$ is irrelevant: 1) $\dot\theta(T_{\rm ws}) < T_{\rm ws}$ in Eq.~(\ref{eq:TRmin_thetadot_TEW}) and  2) $\dot\theta(T_{\rm ws}) < f_a$ in Eq.~(\ref{eq:TRmin_thetadot_fa}).

We now revisit other constraints that are modified because of $S(T_{\rm ws}) > f_a$ and/or $m_a(T_{\rm ws}) < m_a$. The constraint that the axion is in the rotating phase, $\dot\theta(T_{\rm ws}) > m_a(T_{\rm ws})$, appears to be relaxed due to a smaller $m_a(T_{\rm ws})$ than the vacuum value of the mass.
However, a stronger constraint persists when $S(T_{\rm ws}) > f_a$ because the equation of motion of $S$ fixes $\dot\theta^2 = V'(S)/S \simeq m_S^2$ in this regime with $m_S$ the mass of the radial component; in turn, $m_S > m_a$ must hold for the description of a pseudo Nambu-Goldstone boson to apply. Altogether, $\dot\theta(T_{\rm ws}) > m_a$ is hence still necessary (with $m_a$ the vacuum mass), in which case the constraint in Eq.~(\ref{eq:TRmax_thetadot_ma}) applies identically and a period of reheating with $T_R < T_{\rm ws}$ remains necessary as in Sec.~\ref{sec:const_mass}. 

Moreover, even for a theory with suppressed quantum corrections to the potential of $S$, namely supersymmetry, the mass $m_S$ receives a minimum contribution from the quantum correction $\Delta m_S = y_Q m_{\widetilde Q}/4\pi$ that arises from the Yukawa interaction, $y_Q P Q \bar Q$, between the PQ breaking field $P$ and the KSVZ fermions $Q$ and $m_{\widetilde Q}$ is the soft mass for the sfermion $\widetilde Q$. The Yukawa coupling $y_Q$ in principle is constrained experimentally by the mass of $Q$, namely $m_Q = y_Q f_a \gtrsim 1 \TeV$. Requiring $\dot\theta \simeq m_S \ge \Delta m_S$ gives a constraint
\begin{align}
\label{eq:TR_Delta_mSsq}
    T_R \lesssim 8 \GeV 
    \left( \frac{0.1}{c_B} \right)^{ \scalebox{1.01}{$\frac{1}{5}$} }
    \left(\frac{T_{\rm ws}}{130 \GeV} \right)^{ \scalebox{1.01}{$\frac{6}{5}$} } 
    \left( \frac{f_a}{10^4 \GeV} \right)^{ \scalebox{1.01}{$\frac{1}{5}$} }
    \left( \frac{\rm TeV}{m_Q} \right)^{ \scalebox{1.01}{$\frac{1}{5}$} }
    \left( \frac{\rm TeV}{m_{\widetilde Q}} \right)^{ \scalebox{1.01}{$\frac{1}{5}$} }
    \left( \frac{g_*(T_{\rm ws})}{106.75} \right)^{ \scalebox{1.01}{$\frac{1}{5}$} }.
\end{align}
This strong upper bound is in tension with various lower bounds on $T_R$, such as $\dot\theta(T_{\rm ws}) < T_{\rm ws}$ in Eq.~(\ref{eq:TRmin_thetadot_TEW}) and $\dot\theta(T_{\rm ws}) < f_a$ in Eq.~(\ref{eq:TRmin_thetadot_fa}). The experimental lower bound on the Yukawa coupling responsible for the quantum correction, however, is model-dependent. In Appendix~\ref{app:weak_KSVZ_model}, we present an example of a model where the Yukawa coupling therein, $y_P$, can be as small as $10^{-6}$ independently of $f_a$. This leads to a significantly weaker constraint
\begin{align}
\label{eq:TR_Delta_mSsq_relaxed}
    T_R \lesssim 80 \GeV 
    \left( \frac{0.1}{c_B} \right)^{ \scalebox{1.01}{$\frac{1}{5}$} }
    \left(\frac{T_{\rm ws}}{130 \GeV} \right)^{ \scalebox{1.01}{$\frac{6}{5}$} } 
    \left( \frac{10^{-6}}{y_P} \right)^{ \scalebox{1.01}{$\frac{1}{5}$} }
    \left( \frac{\rm TeV}{m_{\widetilde Q}} \right)^{ \scalebox{1.01}{$\frac{1}{5}$} }
    \left( \frac{g_*(T_{\rm ws})}{106.75} \right)^{ \scalebox{1.01}{$\frac{1}{5}$} }.
\end{align}
We will show the viable parameter space for each of the cases in Eqs.~(\ref{eq:TR_Delta_mSsq}) and (\ref{eq:TR_Delta_mSsq_relaxed}). 

The remaining constraints depend also on the field value $S(T_{\rm ws})$ in addition to $T_R$. Therefore, for a given set of model parameters ($m_a, f_a$), one has to examine all of the constraints in the two dimensional parameter space of $S(T_{\rm ws})$ and $T_R$ in order to determine whether ($m_a, f_a$) is viable for explaining the baryon asymmetry.

The parametric resonance rate given in Eq.~(\ref{eq:PReff}) is modified when $S > f_a$. This occurs at the least because the axion mass is suppressed by an amount $m_a \propto S^{-1/2}$ based on Sec.~\ref{subsec:UV_4D}.
Also, in a supersymmetric theory, the potential $V(P)$ becomes increasingly quadratic when $S > f_a$ (see Eq.~\eqref{eq:Leff_2field}) and parametric resonance cannot occur in this limit due to the lack of self-interactions. The motion is highly non-linear since the complex field $P$ is probing the minimum in the radial direction and also the details of the $S$-dependent cosine potential in the angular direction. As a result, we perform a numerical computation of the parametric resonance rate and find the following rate suppressed compared to Eq.~(\ref{eq:PR})
\begin{align}
\label{eq:fund_PR_rate_suppressed}
    %\Gamma_{\rm PR} = \frac{m_a^4}{\dot\theta^3} \left( \frac{f_a}{S} \right)^n.
     \Gamma_{\rm PR} = \frac{m_a^2(S=f_a)}{\dot\theta} \left( \frac{f_a}{S} \right)^n,
\end{align}
with $n\simeq 5$ estimated numerically for $S \lesssim 10 f_a$ in a limited setting as discussed in Appendix~\ref{app:PR_numerics}.
For larger $S$, the rate becomes too suppressed to be determined in a numerically stable manner. Note that the width of the resonance also matters in computing the effective parametric resonance rate.  We take $m_a(S=f_a)$ not much below $\dot{\theta}$ so that the resonance at $S\sim f_a$ is wide ($\delta k/k \sim 1$) and can be easily found numerically. We find that the resonance width remains wide for $S \lesssim 10 f_a$. It is, however, possible that the width becomes narrower at larger $S$. It is also possible that for $m_a(S=f_a) \ll \dot{\theta}$, where the resonance width is already narrow at $S= f_a$, it becomes even narrower for $S > 10 f_a$. We conservatively take the effective parametric resonance rate to be
\begin{align}
\label{eq:PR_rate_suppressed}
 \Gamma_{\rm PR,eff} = \frac{m_a^4(S=f_a)}{\dot\theta^3} \left( \frac{f_a}{S} \right)^n
\end{align} ,
with $n=5$, which is suppressed compared to Eq.~(\ref{eq:PReff}).  We will comment on how larger $n$ affects the viable parameter space at the end of this subsection.
With this large-field suppression, the temperature at which parametric resonance occurs is
\begin{align}
\label{eq:TPR_suppressed}
    T_{\rm PR} \simeq 60 \GeV 
    \left( \frac{T_{\rm ws}}{130 \GeV} \right)^{ \scalebox{1.01}{$\frac{1}{12}$} }
    \left( \frac{c_B}{0.1} \right)^{ \scalebox{1.01}{$\frac{1}{8}$} }
    \left( \frac{T_R}{10 \GeV} \right)^{ \scalebox{1.01}{$\frac{17}{24}$} }
    \left( \frac{m_a}{\rm MeV} \right)^{ \scalebox{1.01}{$\frac{1}{6}$} } 
    \left( \frac{106.75}{g_*(T_{\rm ws})} \right)^{ \scalebox{1.01}{$\frac{7}{48}$} }
    \left( \frac{10 f_a}{S(T_{\rm ws})} \right)^{ \scalebox{1.01}{$\frac{5}{24}$} },
\end{align}
where $S(T_{\rm ws}) > f_a$ delays parametric resonance compared to Eq.~(\ref{eq:TPR}). In particular, $T_{\rm PR} < T_{\rm ws}$ is necessary for the axion rotation to survive until after the electroweak phase transition in order to generate the baryon asymmetry from the charge transfer. The resultant constraint from dangerous electroweak symmetry restoration is estimated in a similar way as Eqs.~(\ref{eq:agg_rate}) and (\ref{eq:rho_rot_TEW}) except for the following two differences. First, the field value is now $S(T_{\rm PR}) = \max(S(T_{\rm ws})(T_{\rm PR}/T_{\rm ws})^4, f_a)$ where $S \propto R^{-3/2} \propto T^4$ during the inflationary reheating era. Second, the scattering rate is enhanced by $(T/k_a)^2$ relative to Eq.~(\ref{eq:agg_rate}) because the radial component fluctuations $\delta S$ are created during parametric resonance and $\delta S$ scatters with the thermal bath without the momentum suppression $k_a$ that originates from the axion derivative coupling.

Once $S(T_{\rm ws})$ is fixed in the exploration of the parameter space, a number of constraints are in order. First, the energy density of the rotation $\dot\theta^2S^2$ should not exceed that of the field reheating the universe as we have assumed. Second, the washout of the rotation may occur when the rotation is approaching the minimum of the radial direction for the following reason. When the rotating field starts to experience the gradient in the angular direction near the minimum, the motion becomes elliptical and thus radial oscillations are induced. The induced radial components get depleted by scattering with the thermal bath. However, the angular gradient continues to induce radial components, and therefore the rotation eventually gets washed out completely. As is shown in Appendix~\ref{app:wo}, this washout via the radial component occurs at a rate given by 
\begin{equation}
\label{eq:wo_S}
    \Gamma_{{\rm wo}, S} \simeq 10^{-5} \frac{T^3}{S^2} \left(\frac{m_a}{m_S} \right)^4,
\end{equation} 
and we impose $\Gamma_{{\rm wo}, S} < H $ for $T > T_{\rm ws}$ so that the rotation survives until the baryon asymmetry production is completed.

\begin{figure}[t!]
    \centering
    \includegraphics[width=0.495\columnwidth]{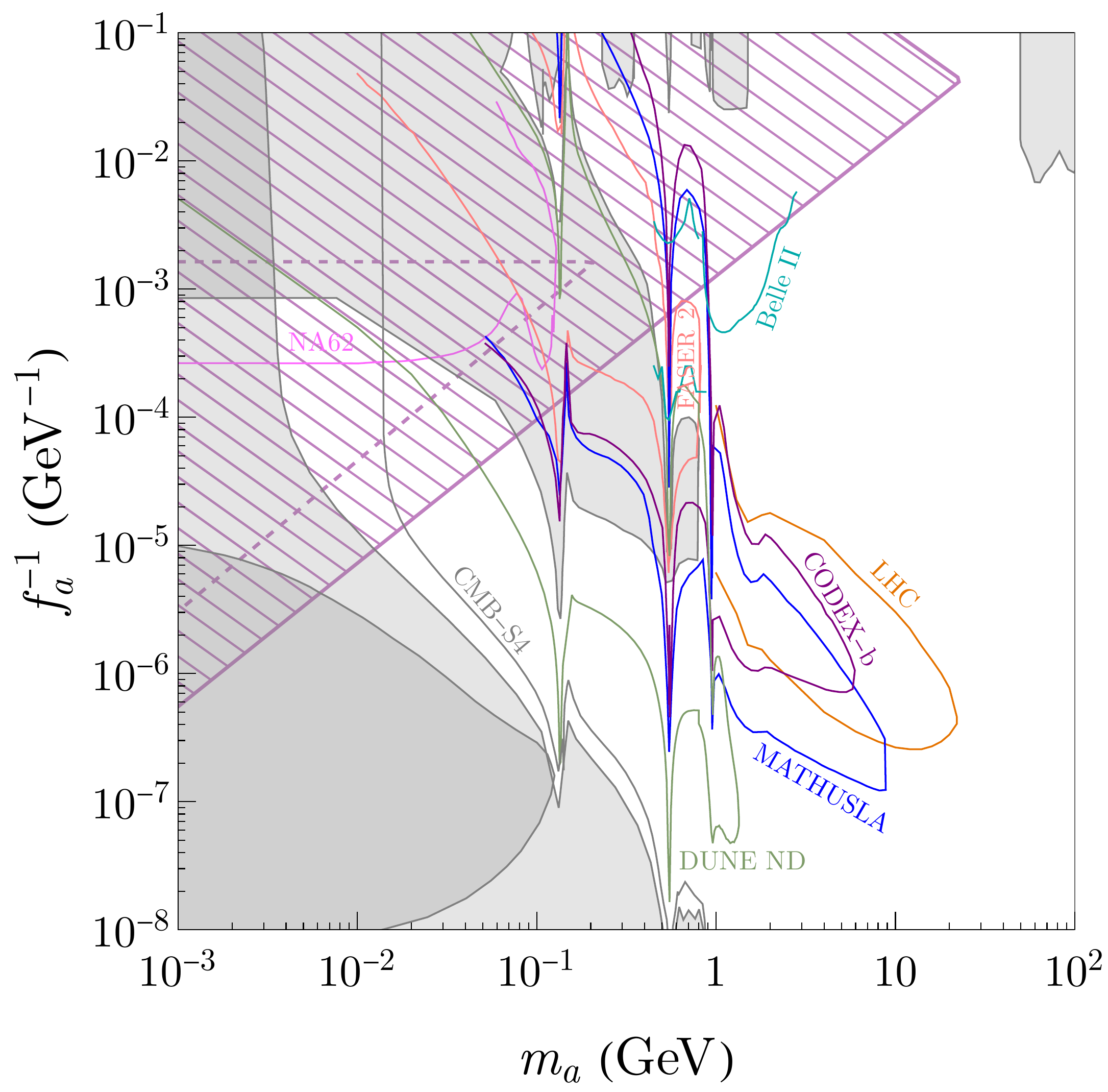}
    \caption{The purple hatched region shows the viable parameter space where the baryon asymmetry can be explained by axiogenesis with a heavy QCD axion, specifically for the case where the axion mass is suppressed at the electroweak phase transition because the radial component has not yet settled to the minimum of the potential.}
    \label{fig:VaryingMass}
\end{figure}

The viable parameter space in the $m_a, f_a$ plane is shown in Fig.~\ref{fig:VaryingMass}. In particular, all of the aforementioned constraints are satisfied in the purple hatched region for some appropriate choices of $S(T_{\rm ws})$ and $T_R$, and thus the observed baryon asymmetry is successfully explained. 

We first discuss the solid boundary of the purple hatched region, which is obtained when we impose Eq.~(\ref{eq:TR_Delta_mSsq_relaxed}) (assuming the model presented in Appendix~\ref{app:weak_KSVZ_model}) rather than Eq.~(\ref{eq:TR_Delta_mSsq}). In this case, the upper right boundary is set by the requirement that the axion is a Nambu-Goldstone boson, $f_a > m_a$, while the lower boundary is set by the consistency between Eq.~(\ref{eq:TRmax_thetadot_ma}) from $\dot\theta(T_{\rm ws}) > m_a$, Eq.~(\ref{eq:rho_rot_TEW}) from evading dangerous electroweak symmetry restoration, and Eq.~(\ref{eq:TPR_suppressed}) from avoiding parametric resonance before the electroweak phase transition. On this boundary, the constraint $T_{\rm PR} \le T_{\rm ws}$ is saturated. If $n>5$ in Eq.~(\ref{eq:PR_rate_suppressed}), which is plausible for $S \gg f_a$, the lower boundary in fact only expands towards large $f_a$ by roughly a factor of three for any $n \ge 6$ since other constraints become more stringent.

We caution that the experimental constraints and prospects in Fig.~\ref{fig:VaryingMass} may change for $f_a < \mathcal{O}({\rm TeV})$ because a model with the QCD anomaly mediated by some of the SM quarks, as presented in Appendix~\ref{app:weak_KSVZ_model}, is necessary to avoid collider constraints on colored particles. The dimension-five operator $a G{\widetilde G} / f_a$ is no longer the correct description of the model, which was assumed in the constraints and prospects. 

Next we comment on the dashed boundary of the purple hatched region, which imposes Eq.~(\ref{eq:TR_Delta_mSsq}) instead of Eq.~(\ref{eq:TR_Delta_mSsq_relaxed}). The upper boundary arises from the conflict between $\dot\theta(T_{\rm ws}) < T_{\rm ws}$ and $\dot\theta(T_{\rm ws}) > \Delta m_S$, whereas the lower boundary originates from the conflict between $S(T_{\rm ws}) > f_a$, $\dot\theta(T_{\rm ws}) > \Delta m_S$, and avoidance of electroweak symmetry restoration. Moreover, the preferred region remains unchanged for $n \ge 6$ because the dominant constraint is no longer $T_{\rm PR} < T_{\rm ws}$.

\subsection{Mirror QCD}
\label{sec:mirror_QCD}
We next discuss the heavy QCD axion arising from the mirror QCD model. The axion mass is suppressed when $T > \Lambda_{\rm QCD}'$, so we may relax the constraint from parametric resonance. This scenario, however, is strongly constrained by the washout of the rotation.

The PQ charge is washed out by the mirror strong sphaleron process. Because of the heavy mirror quark masses, the mirror quark chiral asymmetry no longer suppresses the washout rate, which is given by
\begin{align}
    \Gamma_{\rm wo'} \simeq {\rm min}\left(\alpha_s' \frac{m_u'^2}{T}, 100\,\alpha_s'^{4} T\right) \times \frac{T^2}{S^2}.
\end{align}
Here we assume $\Lambda_{\rm QCD}'<T < v'$, for which the chiral symmetry breaking is dominantly given by the mass rather than the Yukawa coupling. 
We find that no parameter space is allowed after imposing the washout constraint.

However, the washout constraint can be avoided if $\Lambda_{\rm QCD}'$ is sufficiently large. Indeed, for $T < \Lambda_{\rm QCD}'$, the washout from mirror QCD is given by the scattering of the axion with the mirror hadrons, whose rate is exponentially suppressed. We find $\Lambda_{\rm QCD}' \gtrsim 10^6 \GeV \left(10^6 \GeV / S(T_{\rm ws})\right)\left(T_{\rm ws}/130 \GeV\right)^{3/2}$. 
For such a large $\Lambda_{\rm QCD}'$, the constraints on the parameter space are the same as those in the previous two subsections that assume a temperature-independent axion mass. Note that if the KSVZ quark masses exceed $\Lambda_{\rm QCD}'$, the axion mass dependence on $S$ changes and the results in Sec.~\ref{sec:varying_mass} may change.

Unfortunately, the large $\Lambda_{\rm QCD}'$ makes the axion too heavy, terminating the axion rotation before the electroweak phase transition. To have successful axiogenesis, the axion mass must therefore be suppressed by extra approximate chiral symmetry. This can occur in supersymmetric theories where the approximate $R$ symmetry can suppress the axion mass. However, care must be taken so that the possible extra CP phases in supersymmetric theories do not shift the axion potential minimum in the SM QCD by more than $10^{-10}$. We leave the investigation of supersymmetric scenarios to future work.

\subsection{Axion fragmentation at the electroweak phase transition}
\label{sec:PR_EW}

\begin{figure}[t!]
    \centering
    \includegraphics[width=0.495\columnwidth]{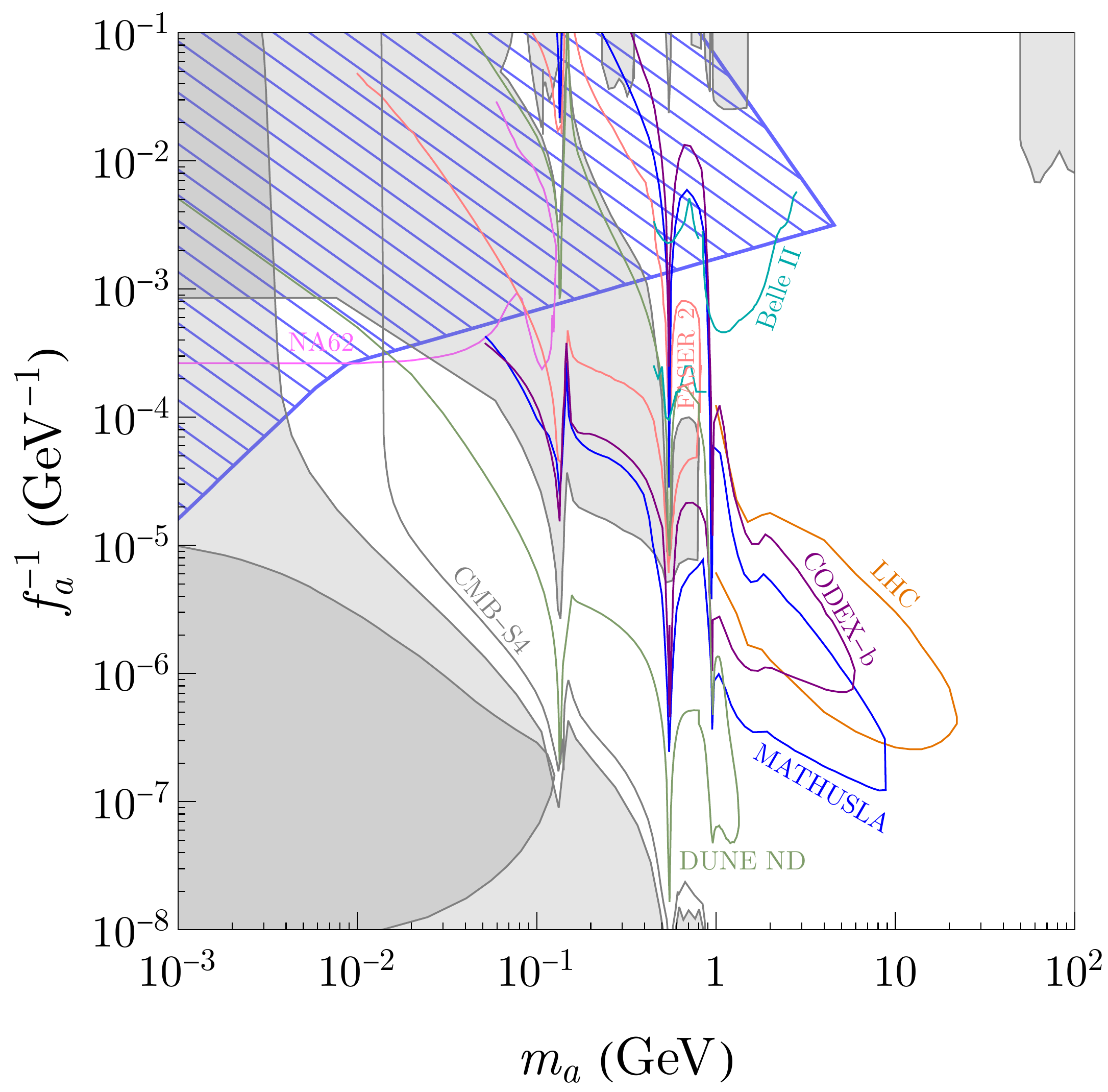}
    \caption{The blue hatched region shows the viable parameter space where the baryon asymmetry can be explained by axiogenesis with a heavy QCD axion, specifically for the tuned case where parametric resonance occurs precisely during the electroweak phase transition.}
    \label{fig:tuned}
\end{figure}

As argued at the beginning of Sec.~\ref{sec:const_mass}, in the absence of entropy production after the electroweak phase transition, the required $\dot\theta(T_{\rm ws}) \simeq 5 \keV$ (from Eq.~(\ref{eq:dtheta_SM})) is too small for the axion to be consistently in the rotating phase in the mass range of interest, $m_a \gtrsim \mathcal{O}({\rm MeV})$. 
To avoid this problem the universe was assumed to undergo a reheating phase during the electroweak phase transition so that entropy production dilutes the baryon asymmetry and the required value of $\dot\theta(T_{\rm ws})$ is accordingly larger. In this section, we consider another possibility where $\dot\theta(T > T_{\rm ws}) \gg m_a$ is exponentially dropping at $T = T_{\rm ws}$ due to parametric resonance so that the instantaneous $\dot\theta(T_{\rm ws})$ is the required value without the need of dilution.

The growth rate of the axion fluctuations from parametric resonance is given in Eq.~(\ref{eq:PReff}). Parametric resonance becomes efficient when both of the following conditions are satisfied. First, the growth rate in Eq.~(\ref{eq:PReff}) needs to be $\gtrsim 10 H$ so that the exponential growth is significantly large. Second, the fundamental parametric resonance rate in Eq.~(\ref{eq:PR}) needs to be larger than the axion-gluon scattering rate in Eq.~(\ref{eq:agg_rate}) so that the scattering processes do not rapidly disturb the Bose enhancement on which parametric resonance relies. Once parametric resonance becomes efficient, the spatial average of $\dot\theta$, denoted by $\vev{\dot\theta}$, drops exponentially quickly, which we have numerically verified. The coherent rotation is then completely destroyed. Therefore, the mechanism considered in this subsection requires that parametric resonance occurs immediately before $T_{\rm ws}$ so that $\vev{\dot\theta}$ decreases to the right amount when the electroweak sphaleron processes fall out of equilibrium at $T_{\rm ws}$.

Requiring parametric resonance to occur at $T_{\rm ws}$ leads to an angular velocity
\begin{align}
\label{eq:thetadot_tune}
    \dot\theta(T_{\rm ws}) \simeq 
    \begin{cases}
    30 \GeV 
    \left( \frac{m_a}{10 \MeV} \right)^{\scalebox{0.9}{$\frac{4}{3}$} }
    \left( \frac{130 \GeV}{T_{\rm ws}} \right)^{\scalebox{0.9}{$\frac{2}{3}$} },
    & m_a \lesssim 20 \MeV 
    \left( \frac{f_a}{10 \TeV} \right)  
    \left( \frac{T_{\rm ws}}{130 \GeV} \right)^{\scalebox{0.9}{$\frac{1}{2}$} }\\ 
    200 \GeV 
    \left( \frac{m_a}{100 \MeV} \right)^{\scalebox{0.9}{$\frac{2}{3}$} }
    \left( \frac{130 \GeV}{T_{\rm ws}} \right)^{\scalebox{0.9}{$\frac{1}{3}$} }
    \left( \frac{f_a}{10 \TeV} \right)^{\scalebox{0.9}{$\frac{2}{3}$} },
    & m_a \gtrsim 20 \MeV 
    \left( \frac{f_a}{10 \TeV} \right)  
    \left( \frac{T_{\rm ws}}{130 \GeV} \right)^{\scalebox{0.9}{$\frac{1}{2}$} }
    \end{cases} ,
\end{align} 
where the upper, lower limits arise from the two aforementioned conditions, respectively. 
The baryon asymmetry is exponentially sensitive to the parameters of theory. Indeed, starting from the value of $\dot{\theta}(T_{\rm ws})$ in Eq.~\eqref{eq:thetadot_tune}, parametric resonance needs to exponentially suppress $\dot{\theta}(T_{\rm ws})$ down to the value in Eq.~\eqref{eq:dtheta_SM} to explain the observed baryon asymmetry of the universe.
Although this requires fine-tuning of the parameters of order $10^{-6}$, the baryon asymmetry is known to be constrained by anthropic requirements~\cite{Tegmark:2005dy}, which may justify the tuning. We therefore analyze this possibility in detail despite the necessity of fine-tuning.

Eq.~\eqref{eq:thetadot_tune} fixes the energy density of the axion fluctuations right after parametric resonance, $\rho_a \simeq k_a^2 f_a^2$, where $k_a \simeq \dot\theta(T_{\rm ws})$ is the typical axion momentum determined by the resonant condition. This energy density must not reheat the universe from axion-gluon scattering and gives the condition
\begin{align}
    \rho_a \times \min\left(1,\frac{\Gamma_{agg}}{H} \right) < \frac{\pi^2}{30} g_*(T_{\rm ws}) T_{\rm ws}^4 .
\end{align} 
The expressions of $\dot\theta(T_{\rm ws})$ in Eq.~(\ref{eq:thetadot_tune}) then lead to an upper bound on the decay constant
\begin{align}
    f_a \lesssim
    \begin{cases}
    3 \TeV 
    \left( \frac{10 \MeV}{m_a} \right)^{\scalebox{0.9}{$\frac{4}{3}$} }
    \left( \frac{T_{\rm ws}}{130 \GeV} \right)^{\scalebox{0.9}{$\frac{8}{3}$} }, 
    & m_a \lesssim 8 \MeV 
    \left( \frac{T_{\rm ws}}{130 \GeV} \right)^{\scalebox{0.9}{$\frac{19}{14}$} }\\ 
    4 \TeV 
    \left( \frac{10 \MeV}{m_a} \right)^{\scalebox{0.9}{$\frac{2}{5}$} }
    \left( \frac{T_{\rm ws}}{130 \GeV} \right)^{\scalebox{0.9}{$\frac{7}{5}$} },
    & m_a \gtrsim 8 \MeV 
    \left( \frac{T_{\rm ws}}{130 \GeV} \right)^{\scalebox{0.9}{$\frac{19}{14}$} }
    \end{cases} ,
\end{align}
of which the two cases determine the lower boundaries of the blue hatched region in Fig.~\ref{fig:tuned}. 
For the small values of $f_a$ constrained here, we find that $\Gamma_{agg} > H$ at $T_{\rm ws}$ so that the axion fluctuations are immediately thermalized after being produced by parametric resonance. We note that for $f_a \gtrsim 6 \times 10^7 \GeV (m_a / 10 \MeV)^{4/3} (130 \GeV / T_{\rm ws})^{7/6}$, axion fluctuations are not immediately thermalized at $T_{\rm ws}$, whose evolution requires a further analysis, but such a large $f_a$ is anyway constrained by SN1987A and also outside the parameter space of interest.

%%%%%%%%%%%%%%%%%%%%%%%%%%%%
\section{Summary and discussion}
\label{sec:summary}
%%%%%%%%%%%%%%%%%%%%%%%%%%%%

In this paper, we have considered the production of the baryon asymmetry of the universe from the rotation of a heavy QCD axion in field space. 
Unlike the usual scenario where the axion is light and long-lived,
the heavy QCD axion is unstable and the dark matter overproduction problem associated with the rotating axion is avoided. This provides a new way to generate the baryon asymmetry and solve the strong CP problem. Nevertheless, the scenario is constrained by various processes, such as the fragmentation of the rotation by parametric resonance, the washout of the rotation by the strong dynamics, and  electroweak symmetry restoration after the rotation disappears. We derived the nontrivial constraints and the associated predictions on the axion mass $m_a$ and the decay constant $f_a$.

Without entropy production after the electroweak phase transition, the required angular velocity of the axion at the electroweak phase transition is $\mathcal{O}(1)$ keV, which is much smaller than the viable heavy QCD axion mass $\gtrsim 1$ MeV. The axion then begins to oscillate around the minimum before the electroweak phase transition and the scenario does not work. Instead, we have considered three viable scenarios.
\begin{enumerate}
    \item 
    Entropy production occurs after the electroweak phase transition (Fig.~\ref{fig:ConstMass}). The required angular velocity is raised and may be much larger than the axion mass. Most of the viable parameter space can be probed by CMB observations and rare $K$ decay searches.
    \item
   The PQ breaking field still rotates on the body of the potential at the electroweak phase transition (Fig.~\ref{fig:VaryingMass}).
   The axion mass is suppressed because of the larger PQ symmetry breaking scale. In the viable parameter space, entropy production is still necessary. Some of the parameter space can be probed by CMB observations and rare $K$ and $B$ decay searches.
   \item
   The rotation fragments into fluctuations during the electroweak phase transition (Fig.~\ref{fig:tuned}). When the sphaleron process goes out of equilibrium, the angular velocity is exponentially suppressed compared to the angular velocity before the fragmentation becomes effective. 
   Almost all of the parameter space can be probed by CMB observations and rare $K$ and $B$ decay searches. 
\end{enumerate}

\begin{figure}[t!]
    \centering
    \includegraphics[width=0.9\columnwidth]{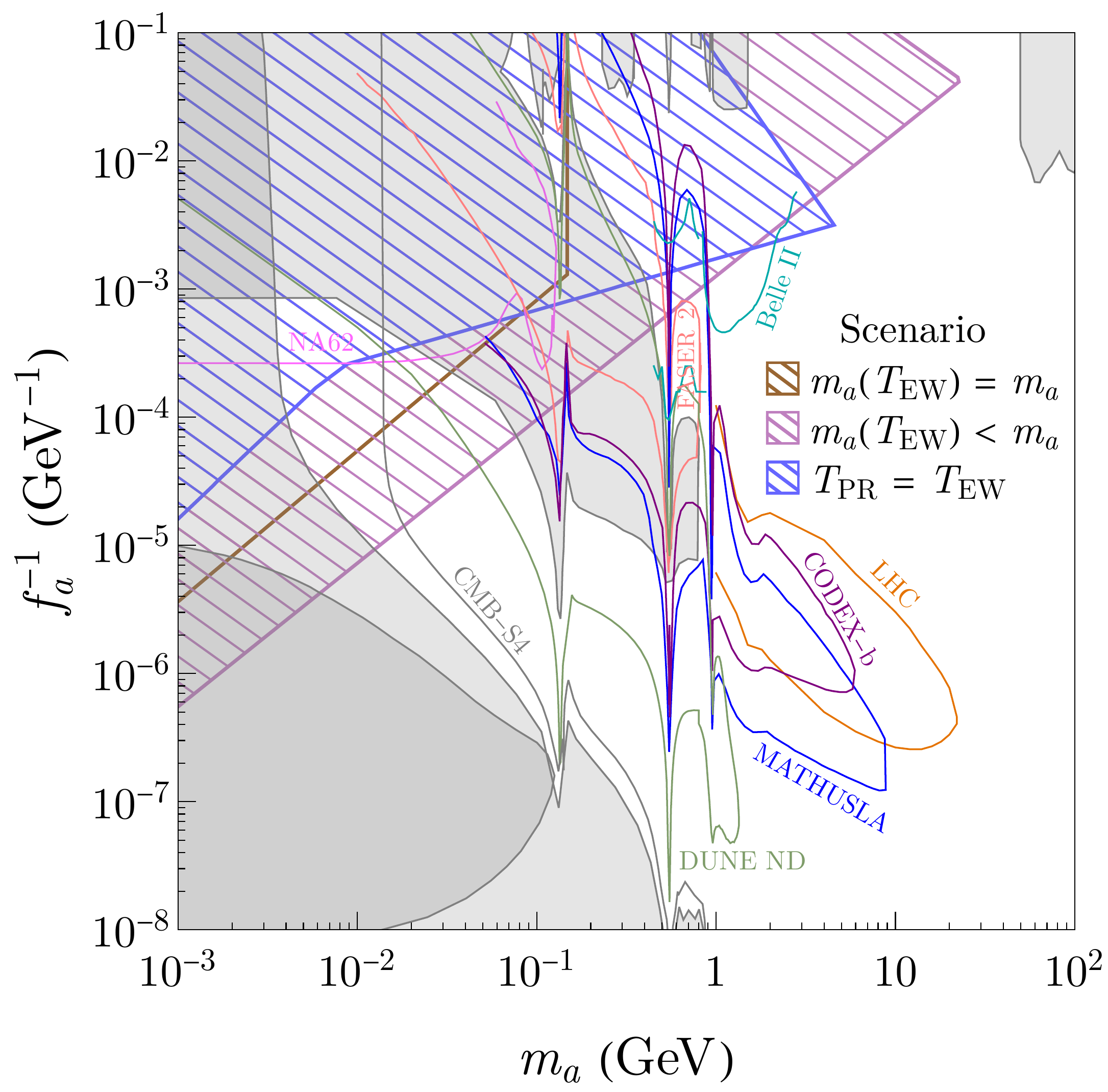}
    \caption{A summary of viable parameter space shown in the hatched regions in which the baryon asymmetry is successfully reproduced, while different colors of hatching correspond to different models and/or cosmological evolution. The brown and blue hatched regions both assume the axion mass is constant, whereas the brown (blue) region considers the case where parametric resonance occurs after (precisely during) the electroweak phase transition at $T_{\rm ws}$. The purple hatched region assumes that $T_{\rm PR} < T_{\rm ws}$ but the axion mass evolves with the radial field value $S$ such that $m_a (T_{\rm ws}) < m_a$ and $S(T_{\rm ws}) > f_a$. The gray shaded regions are excluded by existing constraints, while the regions enclosed by thin boundaries are within the reach of future experimental probes.}
    \label{fig:master}
\end{figure}

In all cases, the decay constant $f_a$ is bounded from above and hence the allowed parameter space can be probed by observations and experiments as summarized in Fig.~\ref{fig:master}. This is because a large $f_a$ means a large energy density of the rotation for a given angular velocity, and the reheating of the universe from the dissipation of the axion rotation tends to wash out the baryon asymmetry by restoring the electroweak symmetry.

Since the heavy QCD axion is cosmologically unstable and cannot be dark matter, a cosmological motivation has been lacking for the regime $f_a\lesssim 10^5$ GeV, even though this part of parameter space is especially interesting for experimental searches. In the paradigm we have presented, heavy QCD axions in the mass range 1 MeV-10 GeV play an important cosmological role in explaining the observed baryon asymmetry via axiogenesis while still solving the strong CP problem. The viable parameter space we have identified in Fig.~\ref{fig:master} can therefore serve as a well-motivated target of experimental searches for heavy QCD axions.

\section*{Acknowledgements}
We thank David Dunsky for providing the data for the CMB constraints and prospects.
The work of R.C. and T.G. is supported in part by the Department of Energy under Grant DE-SC0011842 at the University of Minnesota, and T.G. is also supported by the Simons Foundation.

\appendix

\section{Weakly coupled PQ breaking field in the KSVZ model}
\label{app:weak_KSVZ_model}

As discussed in Sec.~\ref{sec:varying_mass}, the quantum correction from the Yukawa coupling of the PQ symmetry breaking field $P$ to the KSVZ quarks puts a lower bound on the mass of $S$. In this Appendix we present a setup where the Yukawa coupling may be small and therefore reduces the lower bound on the $S$ mass. This model also allows $f_a$ to be much below the mass of new colored particles, thereby realizing $f_a <$ TeV without conflicting with limits from new quark searches at collider experiments.

We introduce vector-like fermions $U$ and $\bar{U}$ where $\bar{U}$ has the same gauge quantum number as the right-handed up quark. The vector-like fermions couple to the right-handed up quark $\bar{u}$, the left-handed quark doublet $Q_u$, the Higgs $H$, and $P$ via the interactions
\begin{align}
    {\cal L} = y_P P \bar{u} U + M_U\bar{U} U + \lambda Q_u \bar{U} H^\dag +{\rm h.c.} .
    \label{eq:ULag}
\end{align}
This structure is enforced by the PQ charge assignment $P(1)$, $\bar{u}(-1)$ with vanishing charges for all other fields.
Assuming that $M_U \gg y_P f_a$, we may integrate out $U\bar{U}$, to obtain the effective Lagrangian
\begin{align}
    {\cal L}_{\rm eff} = \frac{y_P \lambda P}{M_U}Q_u \bar{u} H^\dag + {\rm h.c.} .
\end{align}
The up quark Yukawa coupling ($y_u({\rm TeV})\simeq 6\times 10^{-6}$~\cite{Antusch:2013jca}) is then explained by requiring 
\begin{align}
    y_P \simeq 10^{-6} \left(\frac{M_U}{\rm TeV}\right) \left(\frac{10~{\rm TeV}}{f_a}\right) \left(\frac{1}{\lambda}\right),
\end{align}
where we have assumed a lower bound on the $U$ quark mass of approximately 1 TeV from collider limits.
With this small $y_P$, the quantum correction to the mass of $S$ can therefore be sufficiently suppressed.

However, the setup in general leads to flavor-violating couplings of the axion to up-type quarks. In particular, consider the interactions with the right-handed charm quark $\bar{c}$ and
the left-handed charm quark doublet $Q_c$.
The generic couplings are
\begin{align}
    {\cal L} =  y_P P \bar{u} U + M_U\bar{U} U + \lambda Q_u \bar{U} H^\dag +  \lambda \epsilon_c Q_c \bar{U} H^\dag + y_c Q_c  \bar{c} H^\dag ,
\end{align}
where all couplings $y_P,\lambda, \lambda \epsilon_c,y_c$ can be made real.
The possible mass term $\bar{c}\,U$ and interaction $Q_u \bar{c} H^\dag$ can be removed by rotations of $(\bar{c}, \bar{U})$ and $(Q_c,Q_u)$, respectively. After integrating out $U$ and $\bar{U}$, substituting the VEVs for $P$ and $H$, and removing the axion from the mass terms by the rotation of $\bar{u}$, we obtain
\begin{align}
    {\cal L}_{\rm eff} = \frac{1}{f_a} \partial_\mu a\,\bar{u}^\dag \bar{\sigma}^\mu \bar{u} +
    \begin{pmatrix}
    u & c
    \end{pmatrix}
    \begin{pmatrix}
    y_u & 0   \\
    \epsilon_c y_u & y_c 
    \end{pmatrix}
    \begin{pmatrix}
     \bar{u} \\ \bar{c}
    \end{pmatrix} \frac{v}{\sqrt{2}}.
\end{align}
The diagonalization of the mass term involves a $(\bar{u},\bar{c})$ rotation with an angle $\simeq \epsilon_c y_u/y_c $. Because of the mismatch between the axion-interaction basis and the mass basis, the axion obtains a flavor-violating coupling
\begin{align}
    {\cal L} \supset \epsilon_c \frac{y_u}{y_c} \frac{1}{f_a} \partial_\mu a\,\bar{u}^\dag \bar{\sigma}^\mu \bar{c} + {\rm h.c.} .
\end{align}
This coupling is constrained by rare $D$ meson decays. Using the limit from~\cite{MartinCamalich:2020dfe}, we find
\begin{align}
    \epsilon_c \lesssim 0.05 \left(\frac{f_a}{10^4~{\rm GeV}}\right).
    \label{eq:epslimit}
\end{align}
For $f_a \gtrsim 10^4$ GeV, the upper bound on $\epsilon_c$ is consistent with the naive expectation from the flavor structure needed to explain the Cabibbo angle, which would predict $\epsilon_c \sim 0.1$. On the other hand, $f_a \lesssim 10^4$ GeV requires $\epsilon_c < 0.1$, which can be explained in a flavor model beyond the  Froggatt-Nielsen mechanism with a $U(1)$ symmetry~\cite{Froggatt:1978nt}. Indeed, the Cabibbo angle may arise from the down Yukawa coupling that has $SU(3)_Q\times SU(3)_{\bar{d}}$ flavor charge and the Yukawa coupling of $Q$ with $\bar{U}$ that has $SU(3)_Q\times U(1)_{\bar{U}}$ charge, so we may arrange the flavor symmetry  breaking structure to obtain $\epsilon_c \ll 0.1$.

Note that including the third generation of fermions does not change the result \eqref{eq:epslimit}.
The top quark acquires flavor-violating couplings with the axion and decays into an axion and up/charm quark. The upper bound on the coupling from the rare decay of top quarks is much weaker than the naive expectation from the CKM mixing. All couplings can be made real by rotations of quarks.

Alternatively one may consider a model with vector-like fermions that have the same gauge quantum numbers as the right-handed down-type quarks and mix mainly with the down quark. In this case, a similar analysis shows that the parameter analogous to $\epsilon_c$ is constrained by rare $K$ meson decays and should be $\lesssim10^{-7} (f_a/10^4\,{\rm  GeV})$. Thus for $f_a \sim 10^4$~GeV, the upper bound is substantially smaller than the naive expectation from the Cabibbo angle.

Finally, note that in our setup \eqref{eq:ULag} there is a mass scale $M_U$. One may wonder if this mass scale suppresses the axion mass contribution from UV instanton effects when the instanton scale is larger than $M_U$  (due to the necessity of an $M_U$ insertion). However, when the instanton scale is much above $M_U$, the dominant contribution comes from the diagram without an $M_U$ insertion; $U$ and $\bar{u}$ are attached to $P$, while $Q_u$ and $\bar{U}$ are attached to $H^\dag$. Compared to the usual KSVZ model this leads to an axion mass-squared contribution that is larger by a factor of $M_U/f_a$, and including the effects of the running coupling gives an overall enhancement of $(M_U/f_a)^{1/6}$ to the axion mass provided $M_U> f_a$.

\section{Suppression of parametric resonance for large radial field values}
\label{app:PR_numerics}

The parametric resonance rate for a fixed axion mass is derived in Ref.~\cite{Fonseca:2019ypl} and the results are shown in Eqs.~(\ref{eq:PR}) and (\ref{eq:PReff}) in the absence (presence) of cosmic expansion. In this Appendix, we examine the rate by means of a numerical analysis in the case where the axion mass is a function of the radial field value, which is discussed in Sec.~\ref{sec:models} for the different models and in Sec.~\ref{sec:varying_mass} for the viable parameter space with axiogenesis.

The scenario we analyze in this Appendix assumes that the radial component of the complex field $P$ has a nearly quadratic potential while the angular component receives a mass from explicit PQ breaking that decreases with a large radial field value. To be concrete, we consider a two-field supersymmetric model where the superpotential of the form
\begin{equation}
    W = X (P \bar P - v_{\rm PQ}^2)
\end{equation}
fixes the two fields $P$ and $\bar P$ to the moduli space $P \bar P = v_{\rm PQ}^2$ due to the stabilizer field $X$. The soft masses of $P$ and $\bar P$, written as $V_{\rm soft} = m_P^2 |P|^2 +  m_{\bar{P}}^2 |\bar{P}|^2$, then generate a minimum along the moduli space. In addition, we include an explicit PQ breaking term of the form $V_{\cancel{\rm PQ}} = \mu^3 P + {\rm h.c.}$. Using $\bar P = v_{\rm PQ}^2 / P$, we arrive at the effective Lagrangian
\begin{align}
\label{eq:Leff_2field}
    {\cal L} = \left( 1 + \frac{v_{\rm PQ}^4}{|P|^4} \right) |\partial P|^2 - m_P^2|P|^2 \left( 1 + r_P^2 \frac{v_{\rm PQ}^4}{|P|^4} \right) - \left( \mu^3 P + {\rm h.c.}\right),
\end{align}
where $r_P \equiv m_{\bar{P}}/m_P$.
In the limit $\mu \rightarrow 0$, $|P|$ has a minimum at $\sqrt{r_P} v_{\rm PQ} \equiv v_P$.

\begin{figure}[t!]
    \centering
    \includegraphics[width=0.495\columnwidth]{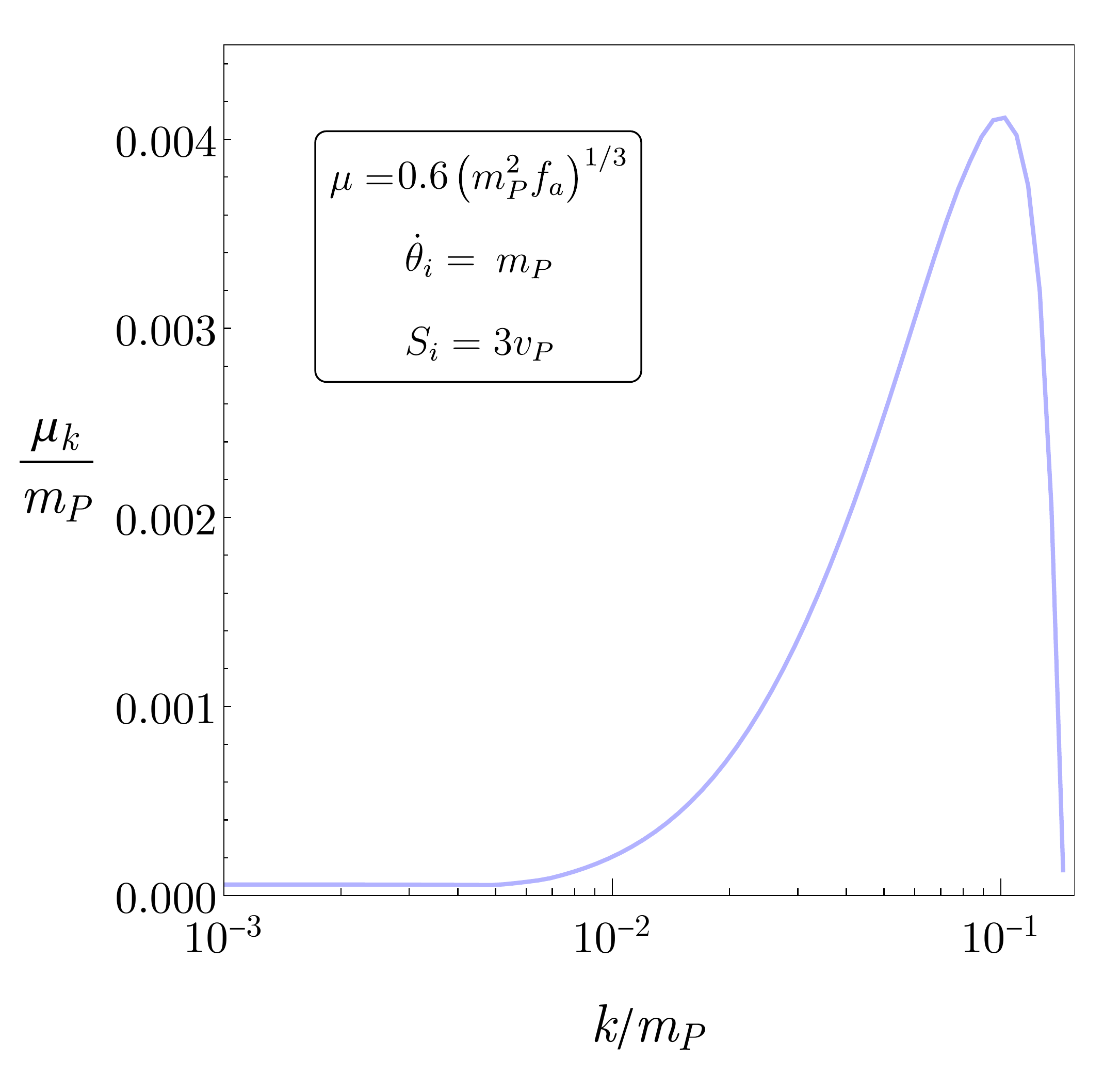}
    \includegraphics[width=0.495\columnwidth]{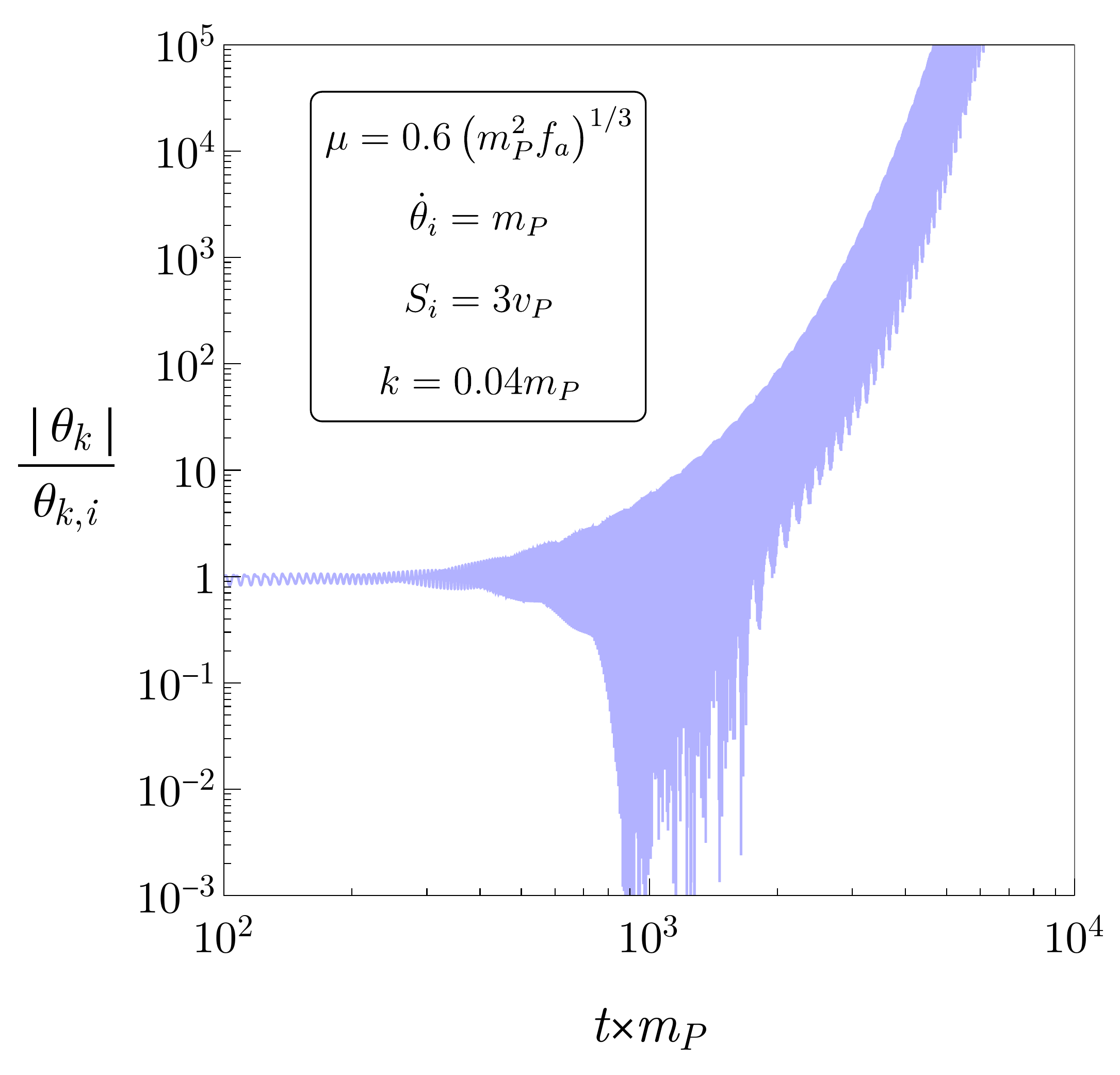}
    \caption{Growth of axion fluctuations due to parametric resonance. The left panel shows the growth rate $\mu_k$ as a function of the momentum mode $k$ both in units of $m_P$. The right panel shows the solution of the fluctuation $\theta_k$ normalized to the initial value $\theta_{k,i}$ as a function of time.}
    \label{fig:PR_fluctuations}
\end{figure}

\begin{figure}[t!]
    \centering
    \includegraphics[width=0.8\columnwidth]{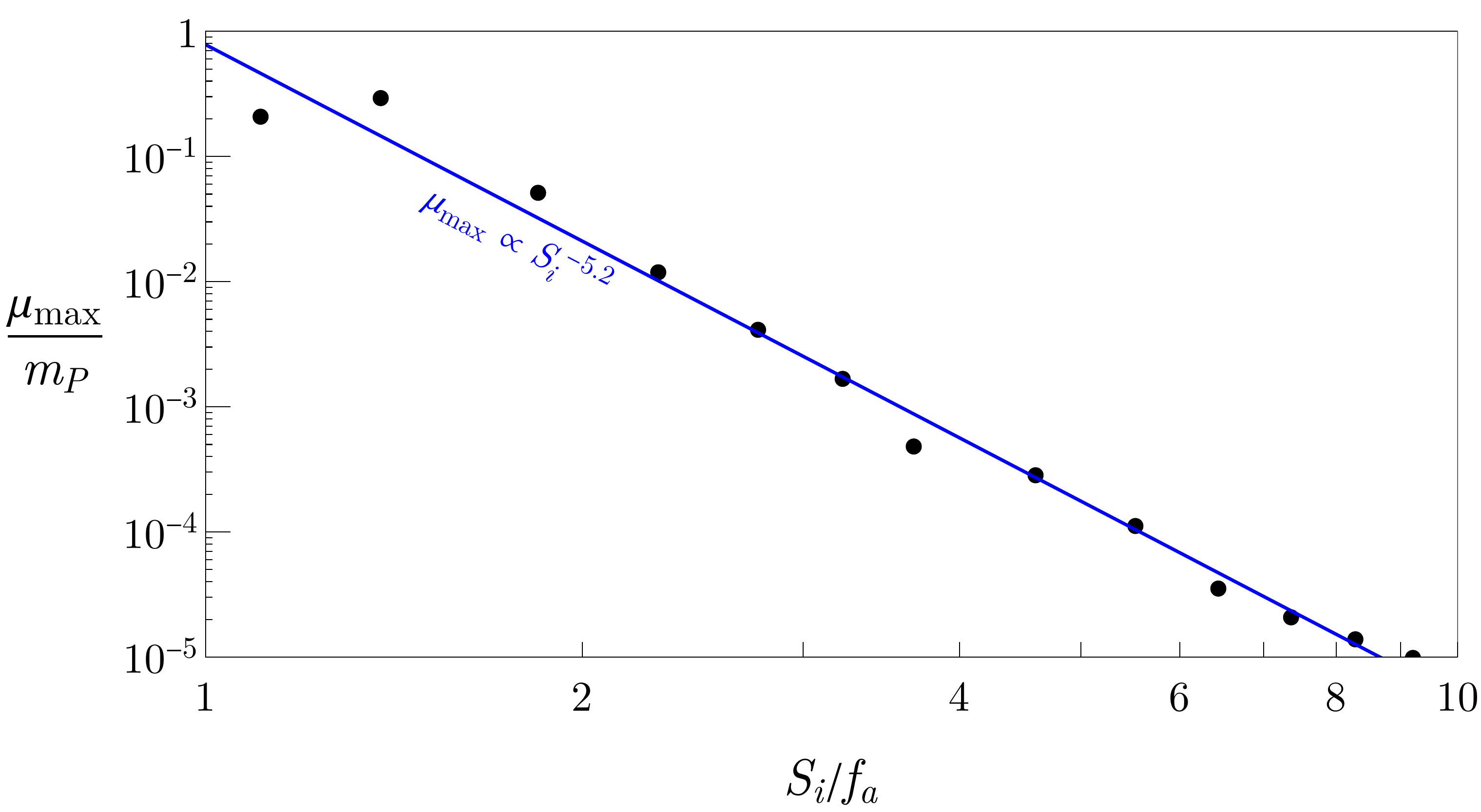}
    \caption{Parametric resonance rate $\mu_{\rm max}$ as a function of the radial field value $S_i$. The black data points obtained numerically suggest a scaling law of $\mu_{\rm max} \propto S_i^n$ with $n \simeq 5$.}
    \label{fig:PR_suppression}
\end{figure}

We apply linear perturbation theory to study the coherent motion and the growing fluctuations. Parametrizing $P$ according to Eq.~(\ref{eq:P}), the equations of motion for the zero modes of the radial component $S_0$ and the angular component $\theta_0$ are given by
\begin{align}
    \ddot S_0 \left( 1 + \frac{v_P^4}{r_P^2 S_0^4} \right) + \dot S_0 \left[ 3 H  \left( 1 + \frac{v_P^4}{r_P^2 S_0^4} \right) - \frac{2 \dot S_0}{S_0} \frac{v_P^4}{r_P^2 S_0^4} \right] \hspace{3.15cm} \nonumber \\ 
        + S_0 \left[ m_P^2 \left( 1 - \frac{v_P^4}{S_0^4} \right) - \dot\theta_0^2 \left( 1 - \frac{v_P^4}{r_P^2 S_0^4} \right) \right] + \sqrt{2} \mu^3 \cos\theta_0 & = 0~, \\
    \ddot\theta_0 \left( 1 + \frac{v_P^4}{r_P^2 S_0^4} \right) + \dot\theta_0 \left[ 3H \left( 1 + \frac{v_P^4}{r_P^2 S_0^4} \right) + \frac{2 \dot S_0}{S_0} \left( 1 - \frac{v_P^4}{r_P^2 S_0^4} \right) \right] - \frac{\sqrt{2} \mu^3 \sin\theta_0}{S_0} & = 0~.
\end{align}
These equations are solved numerically to obtain the zero-mode solutions $S_0(t)$ and $\theta_0(t)$ and then we consider fluctuations, $S(x,t) = S_0(t) + \delta S(x,t)$ and $\theta(x,t) = \theta_0(t) + \delta \theta(x,t)$ around the zero-mode solutions.
We further decompose the fluctuations $\delta S, \delta \theta$ into Fourier modes $S_k$ and $\theta_k$ with momentum $k$.
The linearized equation of motion of the fluctuations of mode $k$ in momentum space is 
\begin{align}
    & \ddot S_k \left[ 1 + \varepsilon \right] + \dot S_k \left[ 3H\left( 1 + \varepsilon \right) - \frac{4 \dot S_0}{S_0} \varepsilon \right] \nonumber \\
         & + S_k \left[ k^2 \left( 1 + \varepsilon \right) + m_P^2 \left( 1 + 3 r_P^2 \varepsilon \right) - \dot\theta^2 \left( 1 + 3 \varepsilon \right) + 2 \varepsilon \left( 5 \left(\frac{\dot S_0}{S_0}\right)^2 - 6 H \frac{\dot S_0}{S_0} - \frac{2 \ddot S_0}{S_0} \right) \right] \nonumber \\
            & \hspace{1 cm} - \dot\theta_k \left[ 2 S_0 \dot\theta_0 \left( 1 + \varepsilon \right) \right] + \theta_k \left[ \sqrt{2} \mu^3 \sin\theta_0 \right] = 0~, \\
    & \ddot\theta_k \left[ 1 + \varepsilon \right] + \dot\theta_k \left[ 3H\left( 1 + \varepsilon \right) + \frac{2 \dot S_0}{S_0} \left( 1 - \varepsilon \right) \right]  + \theta_k \left[ k^2 \left( 1 + \varepsilon \right) - \frac{\sqrt{2} \mu^3}{S_0} \sin\theta_0 \right] + \dot S_k \left[ \frac{2 \dot\theta_0}{S_0} \left( 1 - \varepsilon \right) \right] \nonumber \\
        & + S_k \left[ 2 \frac{\ddot\theta_0}{S_0} \left( 1 - \varepsilon \right) + \frac{\dot\theta_0}{S_0} \left( 6H \left( 1 - \varepsilon \right) + \frac{2 \dot S_0}{S_0} \left( 1 + 3 \varepsilon \right) \right) - \frac{\sqrt{2} \mu^3 \sin\theta_0}{S_0^2} \right] = 0~, 
\end{align}
where we have defined $\varepsilon \equiv v_P^4/r_P^2 S_0^4$.
After obtaining the solutions $S_k$ and $\theta_k$, we consider the time average $\left<\theta_{k,f}\right>$ of $\theta_k$ in the final stage of the numerical solution over a time interval much larger than the oscillation period and compute the growth rate defined as $\mu_k \equiv \log\left(\left<\theta_{k,f}\right>/\theta_{k,i}\right)/\Delta t$ with $\theta_{k,i}$ the initial value.
The growth of fluctuations is shown in Fig.~\ref{fig:PR_fluctuations} for a benchmark point, where the initial condition $\dot\theta_i = m_P$ corresponds to an
initial, approximately circular motion.
In the left panel, we show the growth rate $\mu_k$ as a function of the mode $k$, while in the right panel, the solution of the fluctuation $\theta_k$ is shown for a fixed momentum mode. A maximum rate, $\mu_{\rm max}$ can then be identified among all modes for each given $S_i$. In Fig.~\ref{fig:PR_suppression}, we show the scaling of the maximum growth rate $\mu_{\rm max}$ with the radial field value $S_i$, where the suppression power in Eq.~(\ref{eq:fund_PR_rate_suppressed}) is estimated to be $n \simeq 5$ from the best fit.

\section{Washout by the axion mass and depletion of the radial component }
\label{app:wo}

In this Appendix, we derive the washout rate from the axion mass and the depletion of the radial component when $S \gg f_a$. We consider the following processes. Initially, the rotation is circular. The explicit breaking of the PQ symmetry that generates the axion mass causes deviations from the circular rotation, which induces the radial component. The radial component is then depleted by a rate $\Gamma_S$. We treat the explicit breaking and the depletion as small perturbations.

The potential of the PQ symmetry breaking field at $S \gg f_a$ is approximately given by
\begin{align}
    V(P) = m_S^2|P|^2 + \mu^3 (P + P^\dag),
\end{align}
where the second term is responsible for the axion mass.
We consider small fluctuations $(\alpha(t),\beta(t))$ around the circular motion,
\begin{align}
    P = \frac{1}{\sqrt{2}} S e^{i \theta} = \frac{1}{\sqrt{2}} \bar S(1 + \alpha(t)) e^{i (m_S t + \beta(t)) },
\end{align}
where $\bar S$ is approximately constant. In fact $\bar S$ decreases with a rate $H$, but since we are interested in the case where the washout rate exceeds $H$, we can assume $\bar S$ is a time-independent constant.

The $\alpha$ and $\beta$ equations of motion are, to leading order in $\alpha$, $\beta$, and $\mu^3$,
\begin{align}
    \ddot{\alpha} - 2 m_S \dot{\beta} =& - \sqrt{2}\frac{\mu^3}{\bar S}{\rm cos}(m_St), \nonumber \\
    \ddot{\beta} + 2 m_S \dot{\alpha} =& + \sqrt{2}\frac{\mu^3}{\bar S}{\rm sin}(m_St).
\end{align}
The solution for $\dot{\alpha}$ is given by
\begin{align}
    \dot{\alpha} = \sqrt{2} \frac{\mu^3}{m_S \bar S}{\rm sin}(m_S t),
\end{align}
where we impose $\dot{\alpha}=0$ for $\mu^3 = 0$.
The depletion of the rotational energy density is given~by
\begin{align}
    \dot{\rho} = - \Gamma_S \dot{S}^2 = - \Gamma_S \bar S^2 \dot{\alpha}^2 = - \Gamma_S  \frac{2 \mu^6}{m_S^2}{\rm sin}^2(m_St) \rightarrow - \Gamma_S \frac{\mu^6}{m_S^2},
\end{align}
where the time average has been taken in the last term. Dividing this by the total energy density of the rotation $m_S^2 \bar S^2$, we obtain
\begin{align}
    \Gamma_{{\rm wo}, S} =  \frac{\mu^6}{m_S^4 \bar S^2} \Gamma_S \sim \frac{m_a^4(\bar S)}{m_S^4} \Gamma_S,
\end{align}
where $m_a(\bar S)$ is the mass of the angular component at the large field value $\bar S \gg f_a$. The coupling of the radial component to the gluon gives $\Gamma_S \sim 10^{-5} T^3/\bar S^2$~\cite{Bodeker:2006ij, Laine:2010cq, Mukaida:2012qn} and the final result of $\Gamma_{{\rm wo}, S}$ is used in Eq.~(\ref{eq:wo_S}).

\bibliography{HeavyQCDAxion}

\end{document}